\documentclass[a4paper, 12 pt] {article}
\usepackage{amsmath}
\usepackage{amsfonts}
\usepackage{amsthm}
\usepackage{amscd}

\usepackage{graphicx}
\usepackage{epsfig}
\usepackage{color}
\usepackage{graphics}

\newcommand\zvec{{\bf z}}
\newcommand\qvec{{\bf q}}
\newcommand\pvec{{\bf p}}
\newcommand\rvec{{\bf r}}

\newcommand\Fvec{{\bf F}}

\newcommand\pspace{{\mathbb R}^{2n}}
\newcommand\Rr{{\mathbb R}}
\newcommand\Cc{{\mathbb C}}
\newcommand\Zz{{\mathbb Z}}
\renewcommand\Re{\hbox{\rm Re}\,}
\renewcommand\Im{\hbox{\rm Im}\,}
\newcommand\wn{\hbox{\rm wn}\,}
\newcommand\im{\hbox{\rm im}\,}
\newcommand\Span{\hbox{\rm span}\,}
\newcommand\Sgn{\,\hbox{\rm sgn}\,}

\newcommand\sgn{\,\hbox{\rm sgn}\,}
\newcommand\rank{\hbox{\rm rank}\,}
\newcommand\corank{\hbox{\rm corank}\,}
\newcommand\Tr{\hbox{\rm Tr}\,}
\newcommand\Fvecp{{dF}}
\newcommand\Gvecp{{dG}}

\newcommand\avec{{\bf a}}
\newcommand\bvec{{\bf b}}
\newcommand\cvec{{\bf c}}

\newcommand\uvec{{\bf u}}
\newcommand\vvec{{\bf v}}
\newcommand\wvec{{\bf w}}
\newcommand\svec{{\bf s}}
\newcommand\xvec{{\bf x}}
\newcommand\yvec{{\bf y}}
\newcommand\Zvec{{\bf Z}}

\newcommand\xivec{{\boldsymbol \xi}}
\newcommand\gammavec{{\boldsymbol \gamma}}
\newcommand\etavec{{\boldsymbol \eta}}
\newcommand\thetavec{{\boldsymbol \theta}}
\newcommand\betavec{{\boldsymbol \beta}}
\newcommand\zetavec{{\boldsymbol \zeta}}
\newcommand\chivec{{\boldsymbol \chi}}

\newcommand\Phivec{{\boldsymbol \Phi}}
\newcommand\Psivec{{\boldsymbol \Psi}}
\newcommand\half{\textstyle{\frac12}}
\newcommand\Ord{{O}}

\newcommand\Fmat{{\mathrm F}}
\newcommand\Hmat{{\mathrm H}}
\newcommand\Imat{{\mathrm I}}
\newcommand\Jmat{{\mathrm J}}
\newcommand\Kmat{{\mathrm K}}
\newcommand\Kmatbar{{\overline \Kmat}}
\newcommand\Mmat{{\mathrm M}}
\newcommand\Pmat{{\mathrm P}}
\newcommand\Qmat{{\mathrm Q}}
\newcommand\mmat{{\mathrm m}}
\newcommand\Nmat{{\mathrm N}}
\newcommand\nmat{{\mathrm n}}
\newcommand\Smat{{\mathrm S}}
\newcommand\Tmat{{\mathrm T}}
\newcommand\Umat{{\mathrm U}}
\newcommand\Mp{{d \Mmat}}
\newcommand\Np{{d \Nmat}}
\newcommand\detMp{{d |\Mmat|}}
\newcommand\Cvec{{\bf C}}
\newcommand\Svec{{\bf S}}

\newcommand\Tcal{{\cal T}}

\newcommand\Tt{{T}}
\newcommand\subangle{{\scriptstyle \angle}}

\newcommand\nhat{{\bf \hat n}}
\newcommand\vhat{{\bf \hat v}}
\newcommand\ehat{{\bf \hat e}}
\newcommand\what{{\bf \hat w}}
\newcommand\uhat{{\bf \hat u}}

\newcommand\Avec{{\bf A}}
\newcommand\Bvec{{\bf B}}

\newcommand\Rcal{{\cal R}}

\newtheorem{thm}{Theorem}[section]
\newtheorem{prop}{Proposition}[section]
\newtheorem{defn}{Definition}[section]

\numberwithin{equation}{section}
\begin{document}

\title{The Maslov index and nondegenerate singularities of integrable
  systems}

\author{JA Foxman$^\dag$ and JM Robbins\footnote{E-mail address: {\tt
      j.robbins@bristol.ac.uk}}$^{\dag,\ddag}$\\
$\dag$ School of Mathematics\\ 
University of Bristol,
  University Walk, Bristol BS8 1TW, UK\\
 $\ddag$ The Mathematical
  Sciences Research Institute\\ 1000 Centennial Drive, \#5070,
  Berkeley, CA 94720-5070, USA}

\thispagestyle{empty}

\maketitle
\begin{abstract}
  We consider integrable Hamiltonian systems in $\pspace$ with
  integrals of motion $\Fvec = (F_1,\ldots,F_n)$ in involution.
  Nondegenerate singularities of corank one are critical points of $F$
  where $\rank dF = n-1 $ and which have definite linear stability.
  The set of corank-one nondegenerate singularities is a codimension-two
  symplectic submanifold invariant under the flow.  We show that the
  Maslov index of a closed curve is a sum of contributions $\pm 2$
  from the nondegenerate singularities it encloses, the sign
  depending on the local orientation and stability at the
  singularities.  For one-freedom systems this corresponds to the
  well-known formula for the Poincar\'e index of a closed curve as the
  oriented difference between the number of elliptic and hyperbolic
  fixed points enclosed.  We also obtain a formula for the Liapunov
  exponent of invariant $(n-1)$-dimensional tori in the nondegenerate
  singular set.  Examples include rotationally symmetric $n$-freedom
  Hamiltonians, while an application to the periodic Toda chain is
  described in a companion paper \cite{FR2}.
\end{abstract}

\section{Introduction}\label{sec:intro}

Maslov indices are integers associated with curves on Lagrangian
submanifolds of cartesian phase space $\pspace = \{(\qvec,\pvec)\}$
which count caustics, ie points along the curve where the projection
of the Lagrangian submanifold to the $\qvec$-plane becomes singular.
Caustics are counted with a sign, which is determined by the local
orientation of the curve near the caustic.  For generic curves, the
Maslov index is invariant under deformations which leave the endpoints
fixed.  In semiclassical approximations to the Schr\"odinger equation
(and, more generally, in the short-wave asymptotics of linear wave
equations), Maslov indices appear as $\pi/2$ phase shifts along
classical orbits associated with focusing of families of
orbits.

In classically integrable systems, Maslov indices appear in the EBK
quantisation conditions \cite{keller, maslov}, according to which energy levels
of the corresponding quantum system are given asymptotically by
quantising the actions variables $I_j$ according to
\begin{equation}
  \label{eq:EBK}
  I_j = (n_j + \mu_j/4)\hbar, \quad j = 1,\ldots,n.
\end{equation}
$\mu_j$ is the Maslov index of the closed curve (on a Lagrangian
$n$-torus) traced out by the conjugate angle variable $\theta_j$, and
is even in this case.  For closed curves, Arnold \cite{arnold}  has given a
canonically invariant description of the Maslov index as a
one-dimensional characteristic class, along with an explicit formula
in terms of a winding number in the space of Lagrangian planes.

In the last 20 years there has been much interest in the topology of completely
integrable finite-dimensional Hamiltonian systems.  The generic local
structure and dynamics is given by the Liouville-Arnold theorem 
\cite{arnoldcm},
according to which neighbourhoods of phase space are foliated into
invariant Lagrangian submanifolds diffeomorphic to $\Rr^{n -
  k}\times\Tt^k$, where $\Tt^k$ is the $k$-torus and the dynamics is
linearised by action-angle coordinates.  This local behaviour breaks
down at critical points of the energy-momentum map, where the
dimension of the critical components of the invariant sets drops.  A Morse theory for integrable
Hamiltonian systems, wherein the global topology is described in terms
of these critical sets, has been developed by Fomenko \cite{fomenko},
Eliasson \cite{eliasson}, Vey \cite{vey}
and Tien Zung \cite{tienzung}, among others.  The effect of
singularities on quantum 
wavefunctions are analysed in \cite{toth}.
Obstructions to global action-angle
variables were described by Duistermaat \cite{duistermaat}, and the associated phenomenon
of monodromy has been studied in the classical and quantum contexts
\cite{batescushman, child, vn2001, zhilinskii}.

This paper is also concerned with the topology of completely
integrable systems;  nonzero Maslov indices are manifestations of the
global topology.  We obtain a formula for the Maslov index of a closed
curve in terms of 
the nondegenerate singularities of codimension two (in $\Rr^{2n}$)
enclosed by the curve.  Our result is analogous to the formula for the
Poincar\'e index of closed orbits in planar systems in terms of the
nondegenerate fixed points enclosed by the orbit, and, indeed, for
one-freedom systems coincides with it.

The relationship between the Maslov index and singularities in the
Lagrangian foliation has been studied from a more general point of
view by Suzuki \cite{suzuki}.  Suzuki considers Lagrangian subbundles
of a general symplectic vector bundle, (ie, not just the tangent
bundle of a cotangent bundle), and considers higher Maslov classes of
dimension $4k-3$, though does not consider the case of integrable
Hamiltonian systems.  Trofimov \cite{trofimov}, who also considers
higher Maslov classes, shows that the Maslov indices of Liouville tori
are constant in a connected component of the regular values of the
integrals of motion.

Let us fix
some notations and conventions.
We consider $2n$-dimensional cartesian phase space
$
\Rr^{2n}$, with canonical coordinates $\zvec = (\qvec,\pvec)$, where
$\qvec, \pvec \in \Rr^n$.  The symplectic inner product of $\zvec_1,
\zvec_2 \in \pspace$ is given by
\begin{equation}
  \label{eq:symplectic inner product}
  [\zvec_1,\zvec_2] = \qvec_1\cdot\pvec_2 - 
\pvec_1\cdot\qvec_2 = \zvec_2\cdot \Jmat^{-1} \cdot \zvec_1,
\end{equation}
where
\begin{equation}
  \label{eq:defn of J}
\Jmat = \left(
  \begin{array}{rr}
0&\Imat\\
-\Imat&0
  \end{array}
\right)
\end{equation}
and $\Imat$ is the $n\times n$ identity matrix. 
Given a linear subspace $\lambda \subset \pspace$, denote its
{\it skew-orthogonal complement} by $\lambda^{\subangle}$; this is the subspace
of vectors $\uvec \in \pspace$ for which $[\uvec, \vvec] = 0$ for all
$\vvec \in \lambda$.  $\lambda$ is said to be isotropic if
$\lambda \subset \lambda^\subangle$; in this case $\dim \lambda \le n$.
$\lambda$ is said to be a Lagrangian plane if $\lambda^\subangle =
\lambda$; in this case $\dim \lambda = n$.  The tangent space
$T_{\zvec}\pspace$ is naturally identified with
$\pspace$, in which case the symplectic inner
product~\eqref{eq:symplectic inner product} coincides with the
canonical symplectic form on $T^*\Rr^n \cong \pspace$.

Given smooth functions $F$ and $G$, their Poisson bracket is given by
\begin{equation}
  \label{eq:poisson bracket}
  \{F,G\} = \Fvecp \cdot \Jmat \cdot \Gvecp.
\end{equation}
An integrable system on $\pspace$ is described by a Hamiltonian 
\begin{equation}
  \label{eq:H}
  H = h(\Fvec)
\end{equation}
expressed as a smooth function of $n$ smooth, functionally independent
functions $\Fvec(\zvec) = (F_1(\zvec),\ldots F_n(\zvec))$ in
involution, so that
\begin{equation}
  \label{eq:involution}
  \{F_\alpha,F_\beta\} = 0, \quad 1 \le \alpha,\beta \le n.
\end{equation}
Let
 \begin{equation}
   \label{eq:xivec^alpha}
   \xivec_\alpha = \Jmat\cdot \Fvecp_\alpha.
 \end{equation}
 $\xivec_\alpha$ is the Hamiltonian vector field generated by $F_\alpha$.  
The involution condition \eqref{eq:involution} is equivalent to
 \begin{equation}
   \label{eq:vanishing symplectic inner product}
   [\xivec_\alpha,\xivec_\beta] = 0, \quad 1 \le \alpha,\beta \le n.
 \end{equation}
Let
  \begin{equation}
    \label{eq:lambda}
    \lambda = \Span\{\xivec_1,\ldots,\xivec_n\}.
  \end{equation}
  The condition~\eqref{eq:vanishing symplectic inner product} implies
  that $\lambda(\zvec)$ is isotropic.  If $\xivec_1(\zvec), \ldots,
  \xivec_n(\zvec)$ are linearly independent, then $\lambda(\zvec)$ is
  a Lagrangian plane.
  
  The {\it regular set} of the integrable system, denoted $R$, is the
  set of regular points of $\Fvec$, ie the set on which
  $\Fvecp_1, \ldots, \Fvecp_n$ are linearly independent;
  equivalently, $R$ is the set on which $\lambda$ is Lagrangian. $R$
  is open in $\pspace$.  The complement of $R$, the {\it singular
    set}, denoted $\Sigma$, is the set of critical points of $\Fvec$.
$\Sigma$ is the disjoint union of sets $\Sigma_k$,
  $k = 1,\ldots, n$, on which $d\Fvec$ has corank $k$, 
  \begin{equation}
    \label{eq:Sigma_k}
    \Sigma_k = \{\zvec\, |\, \corank d\Fvec = k\}.
  \end{equation}
  
  Our interest here is in the set $\Sigma_1$, where there is precisely
  one linear relation amongst the $\Fvecp_\alpha$'s.  Given
  $\yvec\in\Sigma_1$, let
\begin{equation}
  \label{eq: linear relation}
  \sum_{\alpha = 1}^n c_\alpha(\yvec) \Fvecp_\alpha(\yvec) = 0,
\end{equation}
where $\cvec(\yvec) \ne 0$.  The relation (\ref{eq: linear relation})
determines $\cvec(\yvec)$ up to a nonzero scalar factor.  We note that
$\cvec(\yvec)$ need not be continuous on $\Sigma_1$.  ((\ref{eq:
  linear relation}) defines line bundles over submanifolds contained
in $\Sigma_1$, which need not be trivial; $\cvec(\yvec)$ defines local
sections of these bundles.)
Given $\yvec \in \Sigma_1$, let
\begin{equation}
  \label{eq:K}
  \Kmat(\yvec) = \sum_{\alpha = 1}^n c_\alpha(\yvec) 
\Jmat {\Fmat_\alpha}''(\yvec),
\end{equation}
where 
\begin{equation}
  \label{eq:Fpp}
  {\Fmat_\alpha}'' = \frac{\partial^2 F_\alpha}{\partial \zvec
  \partial \zvec}
\end{equation}
denotes the Hessian of $F_\alpha$.  Like $\cvec(\yvec)$,
$\Kmat(\yvec)$ is determined up to a nonzero scalar factor, and need
not be continuous on $\Sigma_1$.  $\Kmat(\yvec)$ is an infinitesimal
symplectic matrix, ie 
\begin{equation}
  \label{eq:inf symplectic}
  [\Kmat\cdot\uvec,\vvec] + [\uvec,\Kmat\cdot\vvec] = 0
\end{equation}
for all $\uvec, \vvec \in \pspace$.  Let
\begin{equation}
  \label{eq:tau}
  \tau = \half\Tr\Kmat^2.
\end{equation}
Define the {\it corank-one
  nondegenerate singular set}, denoted $\Delta$, to be the subset of
$\Sigma_1$ on which $\tau \ne 0$;
 \begin{equation}
   \label{eq:Delta}
   \Delta = \{\xvec \in \Sigma_1 \,|\, \tau(\xvec) \ne 0 \}.
 \end{equation}
 Note that, as $\tau$ is determined up to a positive scalar factor,
 $\sgn \tau$ is well defined on $\Delta$. $\tau(\xvec) $ is positive
 (resp.~negative) according to whether $\xvec$ is an elliptic
 (resp.~hyperbolic) fixed point of $\cvec\cdot\Fvec$ (modulo linearly
 neutral directions associated with the action of the remaining
 components of $\Fvec$).

 In Section~\ref{sec: proofs}, we establish that $\Delta$ is an invariant
 codimension-two symplectic submanifold which is invariant under the
 flow of $H$.  While this follows from general results, eg
\cite{tienzung},
we give an independent argument
which serves the developments in subsequent sections.
 In Section~\ref{sec: Maslov formula}, we show that the
 Maslov index of a closed curve $C$ in the regular set is
 typically given by a sum of contributions $\pm 2$ from the
 nondegenerate singular points it encloses, the sign determined by the
 local orientations and signatures $\Sgn \tau$ at the singularities.
 (Typical cases include those for which
 $\Sigma - \Delta$ is contained in a submanifold of codimension
at most three.)  In Section~\ref{sec: transverse stab}, we show that $\Sgn
 \tau$ is related to the linear stability of $\Delta$; given $\xvec
 \in \Delta$ with compact orbit under the integrable flows, we derive
 a formula for the Liapunov exponent, which vanishes or not according
 to whether $\tau(\xvec)$ is positive or negative.  Thus, the signs in
 the Maslov index formula may be regarded as a product of signs
 determining the orientation and stability of the singularities.
 
 Some simple examples are discussed in Section~\ref{sec: examples}.
 For one-freedom systems, our result coincides with the well-known
 formula for the Poincar\'e index of a closed curve $C$ in a planar
 vector field.
 Examples where the formula does not apply, which can arise in
 bifurcations, are also discussed.  In Section \ref{sec:rotations} we consider
 rotationally invariant systems in $\Rr^n$.  In a companion paper, we
 consider a non-separable example, the periodic Toda chain \cite{FR2}.
 
 In what follows, we use the notation $A\cong B$ to indicate that $A$
 and $B$ differ by a nonzero scalar factor (ie, they are projectively
 equivalent).  $A$ and $B$ may be nonzero scalars, vectors, or matrices.

\section{The nondegenerate singular set}\label{sec: proofs}

We show that the nondegenerate singular set $\Delta$ is a
codimension-two symplectic submanifold invariant under the integrable
flow. In outline, the argument is as follows: We introduce a
complex-matrix-valued function, $\Mmat(\zvec)$, whose determinant
vanishes precisely on $\Sigma$.  It is shown that $\Delta$ is an open
subset of the set of regular points of $\det \Mmat =  0$, and a
calculation establishes that the symplectic form is nondegenerate on
$\Delta$.  For brevity, we will denote $\det \Mmat$ by $|\Mmat|$; thus
$\Delta$ is characterised by $|\Mmat| = 0$.

It turns out that the
Maslov index of a closed curve is twice the winding number of $\arg
|\Mmat|$ along the curve.  
This fact, along with the expression (\ref{eq:formula for
  d det M}) for $\detMp$ derived below, is the basis for the Maslov
index formula derived in Section \ref{sec: Maslov formula}.

Let $\Mmat(\zvec)$ be given by
\begin{equation}
  \label{eq:M(z)}
  M_{\alpha \beta}(\zvec) = \frac{\partial F_\beta}{\partial
  p_\alpha}(\zvec) +  i \frac{\partial F_\beta}{\partial q_\alpha}(\zvec).
\end{equation}
The $\alpha^\text{th}$ column of $\Mmat$ has real and imaginary parts
 equal to the
$\qvec$- and $(-\pvec)$-components of $\xivec_\alpha$ respectively.
Then the set of corank-$k$ singularities $\Sigma_k$, 
as given by~(\ref{eq:Sigma_k}), may be
characterised as follows:

\begin{prop}\label{prop: Gamma_k}
  $\zvec \in \Sigma_k \iff \corank \Mmat(\zvec) = k$.
\end{prop}

\begin{proof}
  If $\sum_{\alpha=1}^n a_\alpha \Fvecp_\alpha(\zvec) = 0$ for
  some nonzero $\avec \in \Rr^n$, then it is straightforward to show
  that $\Mmat(\zvec)\cdot \avec = 0$.  Therefore, $ \dim\, \Span \{
    \Fvecp_1(\zvec), \ldots, \Fvecp_n(\zvec)\} \ge
\rank \Mmat(\zvec)$. 
On the other hand, suppose $\avec \in \Cc^n$ is a nonzero right
nullvector of $\Mmat(\zvec)$.  Then $\avec$ is also a right nullvector
of $(\Mmat^\dag \Mmat)(\zvec)$.  The involution condition
\eqref{eq:involution} implies that $(\Mmat^\dag \Mmat)(\zvec)$ is
real.  Therefore, without loss of generality, we may assume that
$\avec$ is real.  For $\avec$ real, it is straightforward to show that
$\Mmat(\zvec)\cdot \avec = 0$ implies that $\sum_{\alpha = 1}^n
a_\alpha \Fvecp_\alpha(\zvec) = 0$.  Therefore, $\rank \Mmat(\zvec)
\ge \dim \Span \{\Fvecp_1(\zvec), \ldots, \Fvecp_n(\zvec)\}$.
\end{proof}
For $\yvec\in\Sigma_1$, the preceding implies that $\cvec(\yvec)$, as
given by \eqref{eq: linear relation}, spans the right nullspace of
$\Mmat(\yvec)$.  Let $\bvec(\yvec)\in\Cc^n$ be the corresponding
(complex) left nullvector, ie
\begin{equation}
  \label{eq:bvec}
  \bvec(\yvec)\cdot\Mmat(\yvec) = 0.
\end{equation}
Like $\cvec(\yvec)$, $\bvec(\yvec)$ is determined up to a (complex)
nonzero scalar factor, and need not be continuous on $\Sigma_1$.

For $\yvec\in\Sigma_1$, let $\betavec(\yvec) \in \pspace$ be given by
\begin{equation}
  \label{eq:betavec}
  \betavec(\yvec) = \left(\Re \bvec(\yvec), \Im \bvec(\yvec)\right).
\end{equation}
Let 
\begin{eqnarray}
  \label{eq:eta and phi}
  \etavec(\yvec) &=& (\Kmat\cdot\betavec)(\yvec),\nonumber\\
  \thetavec(\yvec) &=& (\Kmat\Jmat\cdot\betavec)(\yvec),
\end{eqnarray}
where $\Kmat(\yvec)$ is given by \eqref{eq:K}.  
Then we have the following formula for the derivative of the
determinant of $\Mmat$:
\begin{prop}\label{prop:formula for d det M}
  $\detMp$ vanishes on $\Sigma_k$ for $k > 1$.  For $\yvec \in \Sigma_1$,
\begin{equation} \label{eq:formula for d det M} 
 \detMp(\yvec) \cong \Jmat\cdot\left(\etavec(\yvec) + i \thetavec(\yvec)\right).
\end{equation}
\end{prop}

\begin{proof}
We have the general formula for the derivative of a determinant,
\begin{equation}
  \label{eq:formula for d det M 2}
  \detMp = \Tr \left(\mmat^T\Mp \right),
\end{equation}
where $\mmat(\zvec)$ is the cofactor matrix of $M(\zvec)$.  From
Proposition~\ref{prop: Gamma_k}, if $k > 1$, then 
$\mmat(\zvec)$ vanishes on $\Sigma_k$.
Therefore, $\detMp = 0$ on $\Sigma_k$ for $k > 1$.  

Let $\yvec_0\in \Sigma_1$.  Choose fixed matrices $\Smat_0$ and
$\Tmat_0$ with unit determinant whose first columns are
$\cvec(\yvec_0)$ and $\bvec(\yvec_0)$, respectively.  Let
$\Nmat(\zvec) = \Tmat_0^T \Mmat(\zvec) \Smat_0$. Then $
|\Nmat|(\zvec) = |\Mmat|(\zvec)$, and
\begin{equation}
  \label{eq:prop 2 calculation}
  \detMp = \Tr(\nmat^T\Np),
\end{equation}
where $\nmat(\zvec)$ is the cofactor matrix of $\Nmat(\zvec)$.  By
construction, the
first row and first column of $\Nmat(\yvec_0)$ vanish, so that
$\nmat(\yvec_0)$ has a single nonzero element, namely
$n_{11}(\yvec_0)$.  It follows that
\begin{equation}
  \label{eq:prop 2 calculation cont}
  \detMp(\yvec_0) = n_{11}(\yvec_0){dN_{11}}(\yvec_0) \cong {dN_{11}}(\yvec_0) =
\left(\bvec\cdot \Mp\cdot \cvec\right)(\yvec_0).
\end{equation}
Some straightforward manipulation shows that the preceding is
equivalent to \eqref{eq:formula for d det M} above.
\end{proof}

From Proposition~\ref{prop:formula for d det M}, it follows that the
regular component of the level set $|\Mmat| = 0$, ie, the subset of the
level set on
which $\detMp$ has maximal (real) rank equal to two, is contained in
$\Sigma_1$, and is the set where $\etavec$ and $\thetavec$ are
linearly independent.  Linear independence is implied by the stronger
condition $[\etavec,\thetavec] \ne 0$, which, in turn, is equivalent
to the following condition on $\Kmat$:

\begin{prop}\label{prop: [eta,theta]}
  Let $\yvec\in\Sigma_1$, and let $\ker \Kmat^2(\yvec)$ and $\im
  \Kmat^2(\yvec)$ denote the kernel and image of $\Kmat^2(\yvec)$
  respectively.  If $\tau(\yvec)\ne 0$, then $\dim \im \Kmat^2(\yvec)
  = 2$, $\dim \ker \Kmat^2(\yvec)= 2n-2$, and
  \begin{equation}
    \label{eq:symp decomp}
    \ker \Kmat^2(\yvec) \oplus \im
  \Kmat^2(\yvec) = \pspace
  \end{equation}
is a decomposition of $\pspace$ into symplectic 
 skew-orthogonal subspaces.  Moreover,
  \begin{equation}
    \label{eq:im K^2}
    \im \Kmat^2(\yvec) = \Span\{\etavec(\yvec),\thetavec(\yvec)\},
  \end{equation}
and
\begin{equation}
  \label{eq:sgn[eta,theta]}
  \sgn[\etavec(\yvec),\thetavec(\yvec)] = \Sgn\tau(\yvec).
\end{equation}
\end{prop}
\noindent Proposition~\ref{prop: [eta,theta]} can be deduced from Williamson's
theorem (see, eg, \cite{arnoldcm}). We give an
explicit argument below.  One can also show that, if $\tau(\yvec) = 0$, then
$\Kmat^2(\yvec) = 0$ and $[\etavec(\yvec),\thetavec(\yvec)] = 0$.

\begin{proof}
Let $E_0$ denote the generalised nullspace of $\Kmat(\yvec)$ (the
nullspace of powers of $\Kmat(\yvec)$) and let
$r$ be a positive integer  such that
$\Kmat^r(\yvec)\cdot E_0 = 0$.  Let $E_*
= \im \Kmat^r(\yvec)$.  Then
\begin{equation}
  \label{eq:E_0 + E_*}
  E_0\oplus E_* = \pspace.
\end{equation}
The fact that $\Kmat(\yvec)$ is infinitesimal symplectic implies that $E_0$
and $E_*$ are skew-orthogonal.  For if $\uvec \in E_0$ and $\vvec =
\Kmat^r(\yvec)\cdot \wvec \in E_*$, then 
  \begin{equation}
    \label{eq:E_0 and E_* skew.}
    [\uvec,\vvec] = [\uvec, \Kmat^r(\yvec)\cdot \wvec] = (-1)^r[\Kmat^r(\yvec)\cdot
  \uvec,\wvec] = 0.
  \end{equation}
  As $E_0$ and $E_*$ are complementary and skew-orthogonal, the
  restriction of the symplectic inner product to either is
  non-degenerate, so that both are symplectic subspaces of $\pspace$.

  Next, we show that $\Kmat^2(\yvec)\cdot E_0 = 0$, so that we may
  take $r = 2$ above.  We have that
\begin{equation}
\label{eq: xivec as nullvector}
\Kmat(\yvec)\cdot \xivec_\beta(\yvec) = 0;
\end{equation}
this follows from differentiating $\{F_\alpha,F_\beta\} = 0$,
multiplying the resulting equation by $c_\alpha(\yvec)$ and summing
over $\alpha$.  Thus $\Kmat(\yvec)\cdot \lambda(\yvec) = 0$, so
$\lambda(\yvec)$ is an isotropic subspace of $E_0$.  Since $\yvec\in
\Sigma_1$, $\dim \lambda(\yvec) = n-1$.  As $E_0$ is symplectic, it
follows that $\dim E_0$ is either $2n-2$ or $2n$. 
But the latter
would imply that $\Tr \Kmat^2(\yvec) = 2\tau(\yvec) = 0$, contrary to
assumption.  Thus $\dim E_0 = 2n - 2$ and $\dim E_* = 2$.

The fact that $\dim E_0 = 2n -
2$ implies that $\lambda(\yvec)$ is a maximal isotropic subspace of $E_0$; ie
$\lambda^\subangle(\yvec)\cap E_0 = \lambda(\yvec)$.  It follows that
$\Kmat(\yvec)\cdot E_0 \subset \lambda(\yvec)$.  For if $\uvec\in E_0$
and $\vvec \in \lambda(\yvec)$, then $[\Kmat(\yvec)\cdot \uvec,\vvec]
= -[\uvec, \Kmat(\yvec)\cdot \vvec] = 0$, so that that
$\Kmat(\yvec)\cdot E_0$ is skew-orthogonal to $\lambda(\yvec)$.  Since
$\lambda(\yvec)$ is a maximal isotropic subspace of $E_0$, it follows
that $\Kmat(\yvec)\cdot E_0 \subset \lambda(\yvec)$.  Thus,
$\Kmat^2(\yvec)\cdot E_0 = 0$, so that
\begin{equation}
  \label{eq:E_0 is...}
  E_0 = \ker \Kmat^2(\yvec), \quad E_* = \im \Kmat^2(\yvec).
\end{equation}

Let $\Kmat_*$ denote the restriction of
$\Kmat(\yvec)$ to $E_*$.  From (\ref{eq:E_0 + E_*}),
$\Tr \Kmat_* = \Tr \Kmat(\yvec) = 0$ ($\Kmat$ is necessarily
traceless) and
$\Tr \Kmat_*^2 = \Tr \Kmat^2(\yvec) = 2\tau(\yvec) \ne 0$.  It follows that
the characteristic polynomial of $\Kmat_*$ is given by
\begin{equation}
  \label{eq:char poly of K_*}
  \Kmat^2_* - \tau(\yvec) = 0.
\end{equation}

Let 
\begin{equation}
  \label{eq:beta,Jbeta decomposed}
  \betavec = \betavec_0 + \betavec_*,\quad \Jmat\cdot \betavec = \gammavec_0 + \gammavec_*
\end{equation}
denote the decompositions of $\betavec$ and $\Jmat\cdot\betavec$ into
their respective components in $E_0$ and $E_*$.  From (\ref{eq:bvec})
and (\ref{eq:betavec}), one can verify that $\betavec(\yvec)$ and
$\Jmat \cdot\betavec(\yvec)$ are skew-orthogonal to $\lambda(\yvec)$.
Since $\betavec_*$ and $\gammavec_*$ are necessarily skew-orthogonal to
$\lambda(\yvec)$, it follows that $\betavec_0$ and $\gammavec_0$ are
as well, so that 
\begin{equation}
  \label{eq:beta_0,gamma_0}
  \betavec_0, \gammavec_0 \in E_0 \cap
\lambda^\subangle(\yvec) = \lambda(\yvec).
\end{equation}
Then
\begin{equation}
  \label{eq:beta^2}
  \beta^2 = [\Jmat\cdot\betavec,\betavec] = [\gammavec_*,\betavec_*] + 
[\gammavec_0,\betavec_0] = [\gammavec_*,\betavec_*].
\end{equation}

Now let $\etavec(\yvec) = \Kmat(\yvec)\cdot \betavec(\yvec)$,
$\thetavec(\yvec) = \Kmat(\yvec)\cdot \gammavec(\yvec)$, as in (\ref{eq:eta and phi}).  From
(\ref{eq:beta,Jbeta decomposed}) and (\ref{eq:beta_0,gamma_0}),
\begin{equation}
  \label{eq:redefs of eta,theta}
  \etavec(\yvec) = \Kmat_*\cdot \betavec_*, \quad
\thetavec(\yvec) = \Kmat_*\cdot \gammavec_*.
\end{equation}
Therefore, from (\ref{eq:char poly of K_*}), (\ref{eq:beta^2}) and (\ref{eq:redefs of eta,theta}),
\begin{equation}
  \label{eq:symprod1}
  [\etavec(\yvec),\thetavec(\yvec)] = [\Kmat_*\cdot \betavec_*,
  \Kmat_*\cdot \gammavec_*] = 
-[\Kmat^2_*\cdot \betavec_*,  \gammavec_*] =
-\tau(\yvec)[\betavec_*,  \gammavec_*] = \tau(\yvec)\beta^2,
\end{equation}
from which (\ref{eq:sgn[eta,theta]}) follows.  This in turn implies
 that $\etavec(\yvec)$ and $\thetavec(\yvec)$ are linearly
 independent, so that, from (\ref{eq:redefs of eta,theta}), $\im
 \Kmat^2(\yvec) = E_* = \Span\{\etavec(\yvec),\thetavec(\yvec)\}$.
\end{proof}

Let $\Phivec_t(\zvec)$ denote the flow of the Hamiltonian $H$, and let
$\Smat(\zvec,t)$ denote the linearised flow, ie the $2n$-dimensional
symplectic matrix given by
\begin{equation}
     \label{eq:S_jk}
     \Smat(\zvec,t) = \frac{\partial \Phivec_t}{\partial \zvec}(\zvec).
\end{equation}
$\Smat(\zvec,t)$ satisfies the differential equation
\begin{equation}
  \label{eq:Sdot}
  \dot\Smat(\zvec,t) = \Jmat\Hmat''(\Phivec_t(\zvec))\Smat(\zvec,t).
\end{equation}
The following shows that $\Sigma_1$ is invariant under the flow, while
the quantities $\cvec$ and $\Kmat$ are invariant up to a nonzero
scalar factor.
\begin{prop}\label{prop: c and K invariant}
 If $\yvec\in\Sigma_1$, then $\yvec_t = \Phivec_t(\yvec) \in \Sigma_1$, and
 \begin{eqnarray}
   \label{eq:c(y) cong}
   \cvec(\yvec_t) &\cong& \cvec(\yvec)\\
   \label{eq:K(y) cong}
   \Kmat(\yvec_t) &\cong& \Smat(\yvec,t)\Kmat(\yvec)\Smat^{-1}(\yvec,t).
 \end{eqnarray}
 \end{prop}

\begin{proof}
  Integrability implies that the vector fields $\xivec_\alpha(\zvec)$
  are invariant under the flow, so that
\begin{equation}
   \label{eq:xivec invariant}
   \xivec_\alpha(\Phivec_t(\zvec))
     = \Smat(\zvec,t)\cdot \xivec_\alpha(\zvec).
 \end{equation}
 Since $\Smat(\zvec,t)$ is invertible, $\dim \lambda(\zvec)$ is
 invariant under the flow, which implies that the regular set $R$ and the components
 $\Sigma_k$ of the singular set are separately invariant.  Let
 $\yvec\in \Sigma_1$. (\ref{eq:xivec invariant}) implies that
 $\sum_{\alpha=1}^n c_\alpha(\yvec) \xivec_\alpha(\yvec_t) = 0$. As
 $\yvec_t\in\Sigma_1$, there is just one linear relation amongst the
 $\xivec_\alpha(\yvec_t)$, so (\ref{eq:c(y) cong}) follows.  (\ref{eq:K(y)
   cong}) follows from differentiating \eqref{eq:xivec invariant} to
 get
 \begin{equation}
   \label{eq:diff of xivec invariance}
   \Jmat{\Fmat_\alpha}''(\Phivec_t(\zvec))) \Smat(\zvec,t) =
   \frac{\partial \Smat}{\partial \zvec}(\zvec,t)\cdot
   \xivec_\alpha(\zvec) + \Smat(\zvec,t)\Jmat {\Fmat_\alpha}''(\zvec).
 \end{equation}
 Letting $\zvec = \yvec$ in the above,
 multiplying by $c_\alpha(\yvec)$ and summing over $\alpha$, we get
 \begin{equation}
   \label{eq:Kmat evolve}
   \sum_{\alpha = 1}^n c_\alpha(\yvec) \Jmat 
 {\Fmat_\alpha}''(\yvec_t) = 
 \Smat(\yvec,t)\Kmat(\yvec)\Smat^{-1}(\yvec,t).
 \end{equation}
 From (\ref{eq:K}) and (\ref{eq:c(y) cong}), the left-hand side of
 the preceding is, up to nonzero scalar factor,
 $\Kmat(\yvec_t)$.  (\ref{eq:K(y) cong}) follows.
\end{proof}

\begin{defn}\label{def: Delta}
  The corank-one nondegenerate singular set, denoted $\Delta$, is the
  subset of $\Sigma_1$ on which $\tau \ne 0$.
\end{defn}
 \begin{thm} \label{thm: submanifold}
   $\Delta$ is a codimension-two symplectic submanifold which is
   invariant under the flow.  Given $\xvec \in \Delta$,
   $T_{\xvec}\Delta = \ker \Kmat^2(\xvec)$ and
   $(T_{\xvec}\Delta)^\subangle = \im \Kmat^2(\xvec)$.
 \end{thm}
 \begin{proof} 
   The regular component of the level set $\Sigma_1 = \{\zvec |\,
   |\Mmat|(\zvec) = 0\}$, ie the subset of $\Sigma_1$ on which
   $\detMp$ has maximal rank, is a $(2n-2)$-dimensional submanifold.
   From Propositions~\ref{prop:formula for d det M} and \ref{prop:
     [eta,theta]}, $\Delta$ is contained in the regular component of
   $\Sigma_1$.  Since $\Delta$ is open in $\Sigma_1$, it follows that
   $\Delta$ is itself a $(2n-2)$-dimensional submanifold.  From
   Proposition~\ref{prop:formula for d det M}, the tangent space
   $T_\xvec \Delta$ is the skew-orthogonal complement of
   $\Span\{\etavec(\xvec),\thetavec(\xvec)\}$. From
   Proposition~\ref{prop: [eta,theta]}, it follows that $T_\xvec\Delta
   = \ker \Kmat^2(\xvec)$ and $(T_{\xvec}\Delta)^\subangle = \im
   \Kmat^2(\xvec)$.  From Proposition~\ref{prop: c and K invariant},
   $\Sigma_1$ is invariant under the flow, and for $\yvec\in
   \Sigma_1$, $\tau(\yvec_t)\cong \tau(\yvec)$.  It follows that
   $\Delta$ is invariant.
 \end{proof}

\section{Singularity formula for the Maslov index}\label{sec: Maslov formula}

If $\zvec$ belongs to the regular component $R$ of an integrable
system, then $\lambda(\zvec)$ is a Lagrangian plane.  Arnold \cite{arnold}
showed that $\Lambda(n)$, the space of Lagrangian planes, has
fundamental group $\pi_1(\Lambda(n)) = \Zz$, with continuous closed
curves characterised by an integer winding number.

Let $C$ denote a continuous, oriented closed curve in $R$,
parameterised by $\zvec(s)$, $0\le s \le 1$, with $\zvec(1) =
\zvec(0)$. Then $\lambda(C)$ describes a continuous, oriented closed
curve in $\Lambda(n)$, parameterised by $\lambda(\zvec(s))$.  We
define the Maslov index of $C$, denoted $\mu(C)$, to be the winding
number of $\lambda(C)$ in $\Lambda(n)$, ie
\begin{equation}
  \label{eq:Maslov index}
  \mu(C) = \wn \lambda(C).
\end{equation}
Arnold \cite{arnold} showed that, under certain genericity conditions, for
curves $C$ on a Lagrangian manifold (eg, an invariant torus of an
integrable system), the definition (\ref{eq:Maslov index}) coincides
with the signed count of caustics along $C$
(for curves on invariant tori, this is the Maslov index which enters into the
semiclassical quantisation conditions (\ref{eq:EBK})).

We shall make use of the following explicit formula for the Maslov
index \cite{jmr92}:
 \begin{equation}
   \label{eq:Maslov index formula}
  \mu(C) = \frac{1}{\pi} \Big( \arg |\Mmat|(\zvec(1)) 
 - \arg |\Mmat|(\zvec(0))\Big),
 \end{equation}
 where $\Mmat$ is given by \eqref{eq:M(z)}, and $\arg
 |\Mmat|(\zvec(s))$ is taken to be continuous in $s$ (this is possible
 because, from Proposition~\ref{prop: Gamma_k}, $|\Mmat|$ does not
 vanish in $R$).  Thus, $\mu(C)$ is twice the winding number of the
 phase of $|\Mmat|$ evaluated along $\zvec(s)$.  (The formula
 \eqref{eq:Maslov index formula} may be obtained by averaging the
 count of caustics with respect to a one-parameter family of
 projections. A related formula is given in \cite{litrob87}, and an
 application to resonant tori is discussed in \cite{roblit87}.)

 If $C$ is contractible in $R$, then $\lambda(C)$ is contractible in
 $\Lambda(n)$, so that $\mu(C)$ must vanish.  Therefore, if $C$ has a
 nonzero Maslov index, it must enclose points in the singular set.
 Typically, the Maslov index can be determined by the singularities
 enclosed.  This is most easily demonstrated for a small curve about
 $\xvec \in \Delta$ in the plane spanned by $(T_\xvec \Delta)^\subangle$.
 Let $C^0_\epsilon$ denote the family of curves
 \begin{equation}
   \label{eq:C_epsilon}
   \zvec^0_\epsilon(s) = \xvec + \epsilon(\cos 2\pi s\,
 \etavec(\xvec) + \sin 2\pi s\,
 \thetavec(\xvec)).
 \end{equation}
 For sufficiently small $\epsilon \ne 0$, $C^0_\epsilon$ is contained in
 $R$, and the only singular point it encloses is $\xvec$ (that is,
 $C^0_\epsilon$ can be contracted to $\xvec$ without passing through
 any other singular points).  From Propositions \ref{prop: Gamma_k} and
 \ref{prop:formula for d det M},
 \begin{eqnarray}
   \label{eq:simplest curve}
   |\Mmat|(\zvec^0_\epsilon(s)) &=& |\Mmat|(\xvec) + \epsilon\,
   \detMp(\xvec)\cdot
(\cos 2\pi s\,
 \etavec(\xvec) + \sin 2\pi s\,
 \thetavec(\xvec)) + \Ord(\epsilon^2)\nonumber\\
&=& \text{const} \times i\epsilon [\etavec(\xvec),\thetavec(\xvec)]  e^{2\pi i s} + \Ord(\epsilon^2),
 \end{eqnarray}
 where $\text{const}$ denotes a real, nonzero constant independent of
 $\epsilon$, and $[\etavec(\xvec),\thetavec(\xvec)] \ne 0$ by
 Proposition~\ref{prop: [eta,theta]}.   Thus, 
 \begin{equation}
   \label{eq:arg det M(z)}
   \arg |\Mmat|(\zvec^0_\epsilon(s)) -
   \arg|\Mmat|(\zvec^0_\epsilon(0)) = 2\pi s + \Ord(\epsilon^2).
 \end{equation}
It follows from (\ref{eq:Maslov index formula}) that for sufficiently
 small $\epsilon$,
 \begin{equation}
   \label{eq:my c^0}
   \mu(C^0_\epsilon) = 2.
 \end{equation}
More generally, consider the family of curves $C_\epsilon$ given by
\begin{equation}
  \label{eq:simple curve}
   \zvec_\epsilon(s) = \xvec + \epsilon(\cos 2\pi s\,
 \uvec +  \sin 2\pi s\, \vvec),
\end{equation}
where $\uvec, \vvec \in T_{\xvec}\Rr^{2n}$.  Then
\begin{equation}
   \label{eq:simpler det}
   |\Mmat|(\zvec_\epsilon(s)) = |\Mmat|(\xvec) + \epsilon
   \detMp(\xvec)\cdot
(\cos 2\pi s
 \uvec_* + \sin 2\pi s \vvec_*) + \Ord(\epsilon^2),
 \end{equation}
 where $\uvec_*, \vvec_*$ denote the projections of $\uvec, \vvec$ to
 $(T_{\xvec}\Delta)^\subangle$.  If $\uvec_*$ and $\vvec_*$ are
 linearly independent, then, for sufficiently small $\epsilon$,
 $\mu(C_\epsilon) = +\mu(C^0_\epsilon)$ or  $-\mu(C^0_\epsilon)$ according to whether
 $C_\epsilon$ and $C^0_\epsilon$ are similarly or oppositely oriented,
 ie according to whether $\sgn [\uvec_*,\vvec_*] = \pm
 [\etavec(\xvec),\thetavec(\xvec)]$.  Since $\sgn
 [\etavec(\xvec),\thetavec(\xvec)] = \sgn\tau(\xvec)$ (cf
 Proposition~\ref{prop: [eta,theta]}), we get that
 \begin{equation}
   \label{eq:maslov of C_epsilon}
   \mu(C_\epsilon) = 2 \sgn [\uvec_*,\vvec_*] \sgn \tau(\xvec).
 \end{equation}

 We can extend (\ref{eq:maslov of C_epsilon}) to larger curves in $R$
 as follows.  Let $D^2 = \{(x,y) | x^2 + y^2 \le 1\}$ denote the unit
 two-disk endowed with its standard orientation.
 \begin{defn}\label{def: nondegenerately transverse}
 A {\it nondegenerately transverse disk} is a continuous map $\Svec:
 D^2 \rightarrow \pspace$ smooth on the
 interior of $D^2$ such that  
 i) the image of $\Svec$ is contained in the union $R \cup \Delta$ of 
the regular
     set  and the corank-one nondegenerate singular set,
ii) $\Svec^{-1}(\Delta)$, the preimage of the singular set, is a
     finite set of points $e_j = (x_j,y_j)$ in the interior of $D^2$,
and iii) $\pi_* \circ d\Svec(e_j): T_{e_j}D^2 \rightarrow
     (T_{\xvec_j}\Delta)^\subangle$, where $\xvec_j = \Svec(e_j)$ and 
$\pi_*$ is the
     projection onto $(T_{\xvec_j}\Delta)^\subangle$ with respect to
 the symplectic decomposition $\Rr^{2n} = T_{\xvec_j}\Delta\oplus 
(T_{\xvec_j}\Delta)^\subangle$, is nonsingular.
 \end{defn}
That is, $\Svec$ is a nondegenerately transverse disk if its image
 intersects $\Sigma$ transversally at a finite set of points in
 $\Delta$.  We define $\sigma_j$, the local orientation of $\Svec_j$
 at $\xvec_j$, to be $+1$ (resp.~$-1$) if $\pi_* \circ d\Svec(e_j)$ is
 orientation-preserving (resp.~orientation-reversing), with the
 orientation on $(T_{\xvec_j}\Delta)^\subangle$ determined by the
 restriction of the symplectic form.  Explicitly (cf (\ref{eq:char
   poly of K_*})),
\begin{equation}
   \label{eq:sigma_j}
   \sigma_j = \sgn \left[\uvec_{j*},\vvec_{j*}\right]
            = \sgn \tau(\xvec_j) \,
\sgn \left[\Kmat^2(\xvec_j)\cdot \uvec_j,\vvec_j\right],
 \end{equation}
where
\begin{equation}
  \label{eq:uvec,vvec}
  \uvec_j = \frac{\partial \Svec}{\partial x} (e_j),\quad
 \vvec_j = \frac{\partial \Svec}{\partial y} (e_j),
\end{equation}
and $\uvec_{j*}$, $\vvec_{j*}$ denote the projections of $\uvec$,
$\vvec$ in $(T_{\xvec_j}\Delta)^\subangle$.

\begin{thm}\label{thm: Maslov formula}
  Let $C\subset R$ be the oriented boundary of a nondegenerately
  transverse disk $\Svec$, with $C$ parameterised by $\zvec(s) =
  \Svec(\cos 2\pi s, \sin 2\pi s)$.  Then
\begin{equation}
  \label{eq:maslov formula}
  \mu(C) = 2 \sum_{j=1}^N \sigma_j \Sgn \tau(\xvec_j).
\end{equation}
where the sum is taken over preimages $e_j \in \Svec^{-1}(\Delta)$, and 
$\xvec_j = \Svec(e_j)$.
\end{thm}
If the set $\Sigma - \Delta$ is
contained in a submanifold of codimension three or more, then it can
be shown that any closed curve in $R$ can be realised as the boundary
of some nondegenerately transverse disk (the argument is omitted).  In
this case,  (\ref{eq:maslov formula})
applies to all $C \subset R$.  If
$\Sigma - \Delta$ is of codimension two, then \eqref{eq:maslov
  formula} may not apply. Such cases can arise in connection with
bifurcations, as discussed in Section~\ref{sec: examples}.

Theorem~\ref{thm: Maslov formula} may be summarised as follows: if $C$
encloses only isolated nondegenerate singular points, then its Maslov
index is a sum of contributions $\pm 2$ from each of the singularities
$\xvec_j$ enclosed, the signs depending on the local orientations
$\sigma_j$ and $\Sgn \tau(\xvec_j)$ at the singularity.  In
Section~\ref{sec: transverse stab}, it is shown that if the orbit of
$\xvec_j$ under each of the integrable flows is compact, then $\Sgn
\tau(\xvec_j) = -1$ if these orbits are linearly stable, and $\Sgn
\tau(\xvec_j) = +1$ if they are linearly unstable.  Thus, $\Sgn
\tau(\xvec_j)$ is determined by the stability of the singularity.
We note that if $\Svec_s$ is a continuous family of nondegenerately transverse disks
with fixed boundary $C$,  both the number of singularities
and their local orientations and stabilities may vary with $s$.
However, the signed sum of the products of their orientations and
stabilities remains invariant.

\begin{proof}[Proof of Theorem \ref{thm: Maslov formula}]
Choose $r<1$ so that all of the singular preimages $e_j \in D^2$ are contained
in the disk $x^2 + y^2 \le r^2$.
By continuity, the Maslov index of $C$ is equal to the Maslov index of
the image of this circle under $\Svec$.  Since $\Svec$ is smooth on
the interior of $D^2$, it follows from
(\ref{eq:Maslov
  index formula}) that
\begin{equation}
  \label{eq:mu(C) 1}
  \mu(C) = 
\frac{1}{\pi} \oint_{x^2 + y^2 = r^2} d \arg 
|\Mmat|\circ \Svec.
\end{equation}
By Stokes' theorem, the integration contour can be replaced by a sum
of $N$ positively oriented circles centred at each of the $e_j$'s of radius
$\epsilon$, with $\epsilon$ taken to be small enough so that the circles do not
overlap.  Let $C_j \subset R$ denote the images of these circles under
$\Svec_j$.  Then 
\begin{equation}
  \label{eq:maslov sum over contributions}
  \mu(C) = \sum_{j=1}^N \mu(C_j).
\end{equation}
$C_j$ is parameterised by $\zvec_j(s)$ given by
\begin{equation}
  \label{eq:z_j(s)}
  \zvec_j(s) = \xvec_j +  \epsilon(\cos 2\pi s
 \uvec_j +  \sin 2\pi s \vvec_j) + O(\epsilon^2),
\end{equation}
where $\uvec_j$ and $\vvec_j$ are given by (\ref{eq:uvec,vvec}).  For
$\epsilon$ sufficiently small, the $O(\epsilon^2)$ terms can be
dropped without changing the Maslov index of $C_j$.  Since
$\Svec_j$ is nondegenerately transverse, the projections of $\uvec_j$
and $\vvec_j$ to $(T_{\xvec}\Delta)^\subangle$ are linearly independent.
From (\ref{eq:maslov of C_epsilon}) and (\ref{eq:sigma_j}),
\begin{equation}
  \label{eq:mu(C_j)}
  \mu(C_j) = 2\sigma_j \Sgn\tau(\xvec).
\end{equation}
Substituting into (\ref{eq:maslov sum over contributions}), we get 
the formula (\ref{eq:maslov formula}).
\end{proof}

\section{Transverse linear stability of the nondegenerate singular
  set}\label{sec: transverse stab}

Let $\xvec$ be a nondegenerate corank-one critical point which is
fixed by the flow of $\cvec\cdot \Fvec$.  Clearly, the linear
stability of $\xvec$ under $\cvec\cdot \Fvec$ is determined by the
spectrum of $\Kmat(\xvec) $; in particular, $\xvec$ is elliptic or
hyperbolic according to whether $\tau(\xvec)$ is negative or positive.
Here we consider the linear stability of $\xvec$ under the Hamiltonian
$H =h(\Fvec)$, which need not leave $\xvec$ fixed. We assume that the $\Fvec$-orbit of
$\xvec$ is compact.
One way to address this question would be through a systematic normal form
description of the dynamics in a neighbourhood of the
$\Fvec$-orbit of $\xvec$ (see \cite{tienzung} and, for more detailed
formulations, \cite{cdv-vn} for two degrees of freedom and
\cite{miranda_tienzung} for $n$ degrees of freedom).  Instead, we here
obtain directly an explicit formula for the Liapunov exponent of the
$H$-orbit of  $\xvec$.

Let $\Psivec^\alpha_{s}(\zvec)$ denote the Hamiltonian flow generated
by $F_\alpha(\zvec)$, $\alpha = 1, \ldots, n$.  Integrability implies
that these flows commute.   
Given $\xvec\in \Delta$, let
\begin{equation}
  \label{eq:n flows}
  \xvec_\svec = 
\Psivec^1_{s_1}(\Psivec^2_{s_2}(\cdots\Psivec^n_{s_n}(\xvec))\cdots),
\end{equation}
where $\svec = (s_1,\ldots,s_n)\in \Rr^n$.  Then $\xvec \mapsto
\xvec_\svec$ defines an $\Rr^n$-action on $\Delta$.  Let $\Tcal_\xvec$
denote the orbit of $\xvec$ under this action.  The identity component
of the isotropy subgroup of $\xvec$ is the ray $\svec = s\cvec$
in $\Rr^n$, where $\cvec\cdot d\Fvec(\xvec) = 0$.  Quotienting out by
this subgroup gives an $\Rr^{n-1}$-action on $\Tcal_{\xvec}$.  If
$\Tcal_{\xvec}$ is compact, the Liouville-Arnold theorem implies that
it is topologically an $(n-1)$-dimensional torus, and with
Theorem~\ref{thm: submanifold} that the Hamiltonian flow $\Phivec_t$
generated by $H$ describes a $(2n-2)$-dimensional integrable system in
a $\Delta$-neighbourhood of $\Tcal_{\xvec}$.
In what follows we assume that $\Tcal_{\xvec}$ is compact,
and let $\langle \cdot \rangle_{\Tcal_{\xvec}}$ denote the average over
 $\Tcal_{\xvec}$ with respect to the normalised invariant measure.
The linear stability of $\xvec$ is determined by the (maximal)
Liapunov exponent $\kappa_H(\xvec)$, given by
 \begin{equation}
   \label{eq:Liapunov}
   \kappa_H(\xvec) = \sup_{\chivec \ne 0} 
 \lim_{T\rightarrow \infty} \frac{1}{T} \log
   ||\Smat(\xvec,T)\cdot \chivec||,
 \end{equation}
 where $\Smat$, the linearised flow, is given by (\ref{eq:S_jk}), and $||\cdot||$ denotes
 the Euclidean norm on $\pspace$.  It is a standard result that the
 Liapunov exponent exists for almost all initial conditions and is
 independent of the choice of metric (see, eg, \cite{katokhass}).
 
 For $\xvec_\svec \in \Tcal_{\xvec}$ given by (\ref{eq:n flows}), let
\begin{equation}
  \label{eq:Kmat0}
  \Kmatbar(\xvec_\svec) = \sum_{\alpha = 1}^n c_\alpha \Jmat{\Fmat_\alpha}''(\xvec_\svec).
\end{equation}
 Clearly, $\Kmatbar(\xvec) = \Kmat(\xvec)$, and from 
 Proposition~\ref{prop: c and K invariant} (which holds in particular
 for $H = F_\alpha$, $\alpha = 1, \ldots, n$), it follows that
 \begin{equation}
   \label{eq:K and Kbar}
   \Kmatbar(\xvec_\svec) \cong \Kmat(\xvec_\svec).
 \end{equation}
Unlike $\Kmat$, $\Kmatbar$ is necessarily smooth on 
 $\Tcal_{\xvec}$.
Arguing as in (\ref{eq:diff of xivec invariance}), we have that
\begin{equation}
  \label{eq:kmatbar under flow}
  \Kmatbar(\xvec_\svec) = \Umat(\xvec,\svec)\Kmatbar(\xvec)\Umat^{-1}(\xvec,\svec),
\end{equation}
where
\begin{equation}
  \label{eq:Smat_svec}
  \Umat(\xvec,\svec) = \frac{\partial \xvec_\svec}{\partial
  \xvec}. 
\end{equation}
It follows that $\Kmatbar(\xvec_\svec)$ has a pair of nonzero
eigenvalues $\pm \tau^{1/2}(\xvec)$.  If $\tau(\xvec)> 0$,
Proposition~\ref{prop: [eta,theta]} implies
that
\begin{equation}
  \label{eq:Pmat}
  \Qmat(\xvec_\svec) = \left(\Kmatbar(\xvec_\svec) + \tau^{1/2}(\xvec)\right)\Kmatbar^2(\xvec_\svec)
\end{equation}
is, up to normalisation, a projector onto the one-dimensional
$\tau^{1/2}(\xvec)$-eigenspace of $\Kmatbar(\xvec_\svec)$.
In general, $\Qmat(\xvec_\svec)$ is not symmetric.  The normalised symmetric
projector onto the one-dimensional $\tau^{1/2}(\xvec)$-eigenspace is given by
\begin{equation}
  \label{eq:Qmat}
  \Pmat(\xvec_\svec) = \left(\frac {\Qmat \Qmat^T}{\Tr \Qmat^T \Qmat}\right)(\xvec_\svec).
\end{equation}

\begin{thm}\label{thm:stab}
 Let $\xvec \in \Delta$, and suppose that $\Tcal_{\xvec}$
   is compact.  If $\tau(\xvec) < 0$, then
$\kappa_H(\xvec) = 0$.  
If $\tau(\xvec) > 0$, then
   \begin{equation}
     \label{eq:positive liapunov}
     \kappa_H(\xvec) = \sum_{\alpha=1}^n \left(\frac{\partial h}{\partial
   F_\alpha}(F_1,\ldots, F_n)\right)\!(\xvec)\,\, \kappa_{\alpha}(\xvec),
 \end{equation}
 where $\kappa_{\alpha}(\xvec)$, the Liapunov exponent for $H = F_\alpha$,
 is given by
 \begin{equation}
   \label{eq:kappa^alpha}
   \kappa_{\alpha}(\xvec) = 
 \left | 
 \left \langle
\Tr \left(\Pmat\Jmat {\Fmat_\alpha}'' \right) 
 \right \rangle_{\Tcal_\xvec}
 \right |, 
 \end{equation}
and $\Pmat$ is given by (\ref{eq:Qmat}).
 The Liapunov exponents $\kappa_{\alpha}(\xvec)$ do not all vanish;
 in particular,
 \begin{equation}
   \label{eq:sum of kappaalphas}
   \sum_{\alpha=1}^n c_{\alpha}(\xvec) \kappa_{\alpha}(\xvec) = 
 \tau^{1/2}(\xvec).
 \end{equation}
 \end{thm}
 We note that the transpose $\Qmat^T$ in (\ref{eq:Qmat}) is defined with respect to the
 Euclidean inner product. For a general non-Euclidean metric one would a
 obtain a more general expression for $\kappa_{\alpha}$; however, its
 value would be unchanged.

\begin{proof}
  
  As noted above, in a neighbourhood of $\Tcal_\xvec$, the restriction of the flow of
  $H$ to $\Delta$, regarded as a symplectic manifold of dimension
  $2n-2$, is integrable.  As Liapunov exponents for compact integrable flows
  vanish, the right-hand side of \eqref{eq:Liapunov} vanishes for
  $\chivec$ tangent to $\Delta$, so we may restrict $\chivec$ 
to the two-dimensional transverse plane
  $E_*(\xvec)$ defined in (\ref{eq:E_0 is...}).  As nonzero Liapunov exponents for Hamiltonian systems
  occur in signed pairs, for any nonzero $\chivec \in
  E_*(\xvec)$, the limit on the right-hand side of \eqref{eq:Liapunov}
  either vanishes, or else is equal to $\pm \kappa_H(\xvec)$.
  Therefore, if one takes the absolute value of the expression on the
  rhs of
  \eqref{eq:Liapunov}, the supremum over $\chivec$ is no longer
  necessary.
  
  We consider first the case $\tau(\xvec) = -\omega^2 < 0$.  For
  $\xvec_\svec\in\Tcal_\xvec$, $\Kmatbar(\xvec_\svec)$ has a pair of
  imaginary eigenvalues $\pm i\omega$.  Let
  $\zetavec(\xvec_\svec),\zetavec^*(\xvec_\svec)$ denote corresponding
  conjugate eigenvectors.  The real and imaginary parts of
  $\zetavec(\xvec_\svec)$ span $E_*(\xvec_\svec)$.  Therefore,
  $\left[\zetavec(\xvec_\svec),\zetavec^*(\xvec_\svec)\right]$ cannot
  vanish.  The normalisation condition
\begin{equation}\label{eq:etavecnorm}
  \left[\zetavec(\xvec_\svec),\zetavec^*(\xvec_\svec)\right] = i
\end{equation}
determines $\zetavec(\xvec_\svec)$ up to a complex phase factor (we
do not assume that this phase factor can be chosen to make
$\zetavec(\xvec_\svec)$ continuous on $\Tcal_\xvec$).  

Let $\xvec_t = \Phivec_t(\xvec)$.  \eqref{eq:kmatbar under flow} implies that
$\Smat(\xvec,t)\cdot\zetavec(\xvec)$ is proportional to
$\zetavec(\xvec_t)$.  
Since the symplectic inner product is preserved
under the linearised flow, the normalisation condition
\eqref{eq:etavecnorm} implies that $\Smat(\xvec,t)\cdot\zetavec(\xvec)$
differs from $\zetavec(\xvec_t)$ by a phase factor, so that
\begin{equation}\label{eq:complex norm}
  ||\Smat(\xvec,t)\cdot\zetavec(\xvec)|| = ||\zetavec(\xvec_t)||,
\end{equation}
where $||\zetavec||^2 = ||\Re\zetavec||^2 + ||\Im\zetavec||^2$.  Since
$||\zetavec(\xvec_\svec)||$ is bounded on $\Tcal_\xvec$, 
$||\Svec(\xvec,t)\cdot \zetavec(\xvec)||$ is bounded in $t$, so that
$\kappa_H(\xvec) = 0$.

Next we consider the case $\tau(\xvec) > 0$.  For convenience, we
first assume that $H = F_\alpha$.
Take $\chivec \in E_*(\xvec)$ to be an eigenvector of
$\Kmatbar(\xvec)$ with eigenvalue $\tau^{1/2}(\xvec)$.
We may write \eqref{eq:Liapunov} as
\begin{equation}
  \label{eq:time integral}
  \kappa_{\alpha}(\xvec) = \left|\lim_{T\rightarrow\infty} \frac{1}{T}
  \int_0^T
\frac{d}{dt} \log ||\chivec(t)|| \, dt\right|
\end{equation}
where $\chivec(t) = \Smat(\xvec,t)\cdot \chivec$.
From
\begin{equation}
  \label{eq:d/dt S}
  \frac{d}{dt} \Smat(\xvec,t) = \Jmat{\Fmat_\alpha}''(\xvec_t)
\Smat(\xvec,t)
\end{equation}
it follows that
\begin{equation}
  \label{eq:d/dt log w}
  \frac{d}{dt} \log ||\chivec(t)|| = 
\frac{\chivec(t)\cdot \Jmat{\Fmat_\alpha}''(\xvec_t)\cdot\chivec(t)}
{\chivec(t)\cdot\chivec(t)} = 
\Tr \left(\Pmat(\xvec_t) \Jmat {\Fmat_\alpha}''(\xvec_t)\right),
\end{equation}
where $\Pmat$ is given by (\ref{eq:Qmat}).
The fact that the commutative flow (\ref{eq:n flows}) is transitive on
$\Tcal_\xvec$ implies that the Liapunov exponent is constant on 
$\Tcal_\xvec$,
and therefore is equal to its average. (\ref{eq:time integral}) and (\ref{eq:d/dt log w}),
\begin{equation}
  \label{eq:averaging 1}
  \kappa_{\alpha}(\xvec) = \left\langle 
\kappa_{\alpha}(\xvec_\svec)
\right\rangle_{\Tcal_\xvec} = 
\left| \left \langle
\lim_{T\rightarrow \infty}
\frac{1}{T}
\int_0^T
\Tr 
\left(\Pmat((\xvec_\svec)_t) \Jmat {\Fmat_\alpha}''((\xvec_\svec)_t)\right )
\right| \right \rangle_{\Tcal_\xvec}
.
\end{equation}
If the integrand in (\ref{eq:averaging 1}) is averaged over
$\Tcal_\xvec$, the average over $t$ becomes redundant.
We obtain
\begin{equation}
  \label{eq:kappa final}
  \kappa_{\alpha}(\xvec) = \left|
\left\langle 
\Tr \Pmat \Jmat{\Fmat_\alpha}''
\right\rangle_{\Tcal_\xvec}\right|.
\end{equation}

For general $H = h(F_1,\ldots,F_n)$, the expression for the Liapunov
exponent is obtained by replacing $\Jmat{\Fmat_\alpha}''$ by
$\Jmat\Hmat''$ in the preceding calculation.  We have that
\begin{equation}
  \label{eq:JH'' w}
  \chivec \cdot \Jmat \Hmat''\cdot \chivec = \sum_{\alpha = 1}^n
  \frac{\partial h}{\partial F_\alpha} \chivec \cdot\Jmat{\Fmat_\alpha}'' \cdot \chivec +
\sum_{\alpha,\beta = 1}^n \frac{\partial^2 h}{\partial F_\alpha
\partial  F_\beta} (\chivec\cdot \xivec_\alpha) \left[\xivec_\beta,\chivec\right ].
\end{equation}
For $\chivec \in E_*(\xvec)$, the second term vanishes (cf
Proposition~\ref{prop: [eta,theta]}).  
Thus, we get
\begin{equation}
  \label{eq:kappa_H}
  \kappa_H(\xvec) = 
\sum_{\alpha=1}^n
\left(\frac{\partial h}{\partial F_\alpha}\right)\!(\xvec)\,\, \kappa_{\alpha}(\xvec),
\end{equation}
in accord with (\ref{eq:positive liapunov}).
If we let  $H(\zvec) = \sum_{\alpha = 1}^n c_\alpha F_\alpha(\zvec)$ and
use the fact that $\Pmat\sum_{\alpha=1}^n c_\alpha\Jmat
\Fmat_\alpha''= \Pmat \Kmatbar = \tau^{1/2}\Pmat$, we get from
\eqref{eq:kappa_H}
that
\begin{equation}
  \label{eq:sum rule}
  \sum_{\alpha = 1}^n c_{\alpha}(\xvec)\kappa_{\alpha}(\xvec) = \tau^{1/2}(\xvec),
\end{equation}
as in (\ref{eq:sum of kappaalphas})
\end{proof}

\section{Examples}\label{sec: examples}
\subsection{One-freedom systems.}\label{sec: one dim}
Let $H(q,p)$ be a smooth Hamiltonian on $\Rr^2$.  The
 singular set $\Sigma$ consists of critical points of $H$, ie fixed
 points of the flow, and the nondegenerate singular set $\Delta$
 consists of isolated critical points where $ \Tr (\Jmat \Hmat'')^2 =
 -2 \det (\Jmat \Hmat'') \ne 0$. These are the hyperbolic ($\det \Jmat
 \Hmat'' < 0$) and elliptic ($\det \Jmat\Hmat'' > 0$) fixed points of $H$.
 The space of Lagrangian planes $\Lambda(1)$ is just the projective
 line $RP^1$, and the Maslov index \eqref{eq:Maslov index} of a closed oriented
 curve in $R = \Rr^2 - \Sigma$ is just ($-2$ times) the Poincar\'e
 index (see, eg, \cite{guho}) of
 the velocity field $\Jmat dH$ around the curve.  If all the fixed
 points are isolated and nondegenerate (ie, $\Sigma = \Delta$), then
 the result \eqref{eq:maslov formula} is equivalent to the standard
 expression for the Poincar\'e index as the number of elliptic fixed
 points minus the number of hyperbolic fixed points enclosed by a
(positively oriented) curve.

The Hamiltonian $H_0 = p^2/2 - q^3/3$ has a single degenerate fixed
point at the origin.  Formula \eqref{eq:maslov formula} does not apply
in this case, but it is straightforward to show that the Poincar\'e
index (and Maslov index) vanishes for every closed curve in the plane,
whether or not it encloses the origin. The Hamiltonian $H_0$ can be
embedded in a one-parameter family $H_a = p^2/2 - (q^3/3 +a q)$ which
undergoes a saddle-centre bifurcation at $a = 0$, and for which
\eqref{eq:maslov formula} applies for $a\ne 0$.  The index of a closed
curve $C$ about the origin is constant through the bifurcation,
vanishing for $a \rightarrow 0-$ because there are no fixed points,
and for $a \rightarrow 0+$ because the contributions from the saddle
at $-\sqrt{a}$ and the centre at $\sqrt{a}$ cancel.  A two-dimensional
example is given next.

\subsection{Two-freedom bifurcation.}\label{sec: bifurcation}

\begin{figure}
\begin{center}
\input{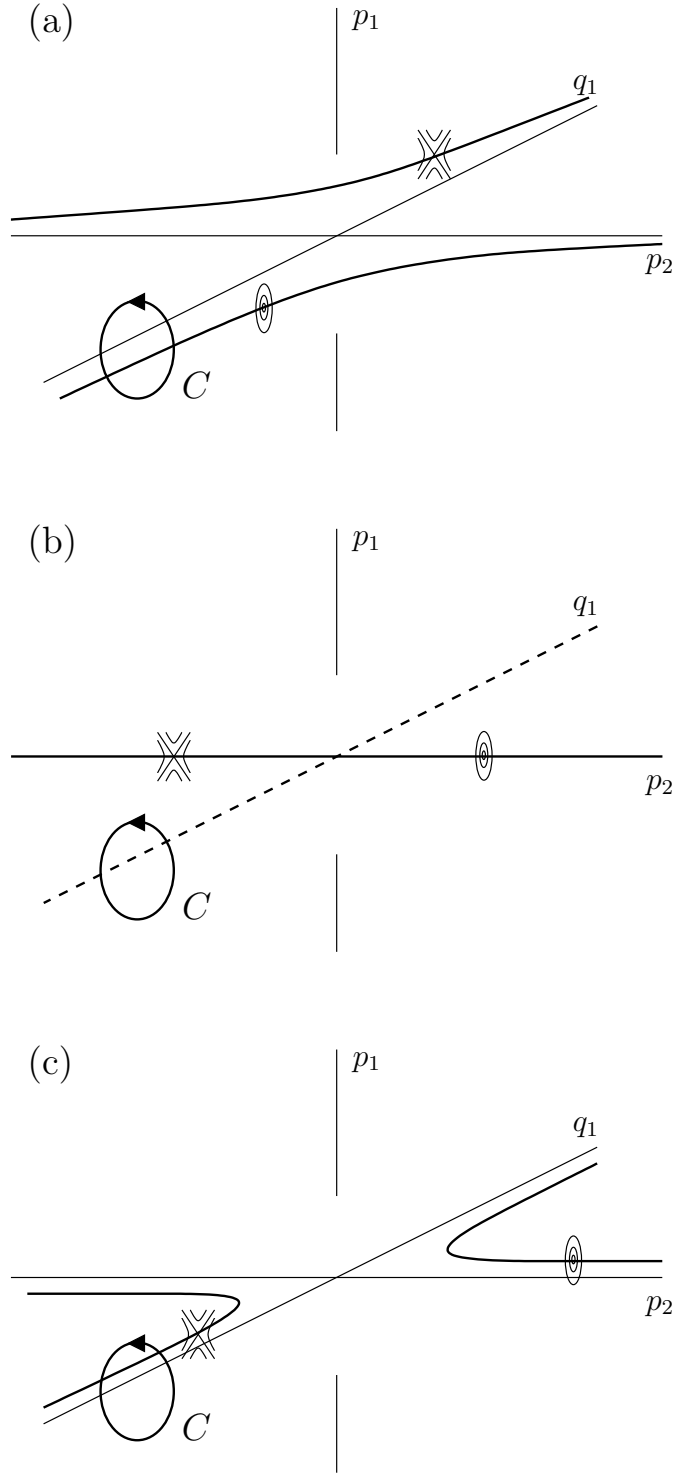}
\caption{Cycle in phase space with
  $H = \half p_1^2 + \half p_2 q_1^2 - \epsilon q_1$.  $\mu(C) =
  2$. (a) For $\epsilon < 0$, $C$ encloses a stable singularity with
  positive orientation.  (b) For $\epsilon = 0$, $C$ encloses a
  degenerate singularity.  (c) For $\epsilon > 0$, 
$C$ encloses an unstable singularity with
  negative orientation.}
\label{fig: bifur}
\end{center}
\end{figure}

The integrable system
\begin{equation}
  \label{eq: eps = 0}
  F_1 = \half p_1^2 + \half p_2 q_1^2, \quad F_2 = p_2,
\end{equation}
for which $q_2$ is an ignorable coordinate, is a nongeneric example
where the degenerate singular set $\Sigma - \Delta$ is of codimension two.  The singular
set is given by $p_1 = 0, p_2 q_1 = 0$.  With  appropriate choices
of
$\cvec$, 
$\Tr \Kmat^2 = -p_2$.  The
nondegenerate singular set $\Delta$ has two disconnected components,
namely $p_1 = q_1 =0, p_2 > 0$ (stable fixed point in the
$(q_1,p_1)$-plane) and $p_1 = q_1 =0, p_2 < 0$ (unstable fixed point
in the $(q_1,p_1)$-plane.)  The degenerate singular set is the
coordinate plane $\pvec = 0$.  The closed curve $C$, with $\qvec$
fixed and $\pvec = \epsilon(\cos\theta,\sin\theta)$, encloses only the
degenerate singular set (see Figure~\ref{fig: bifur}(b)).  A calculation, for example using
\eqref{eq:Maslov index formula}, shows that $\mu(C) = 2$.

The degeneracy can be lifted by embedding the system in a
one-parameter family, 
\begin{equation}
  \label{eq: eps ne 0}
  F_1^\epsilon = \half p_1^2 + \half p_2 q_1^2 - \epsilon q_1, \quad F_2 = p_2.
\end{equation}
The singular set is given by $p_1 = 0, p_2 q_1 = \epsilon$, and $\Tr
\Kmat^2 = -p_2$ as before.  The system undergoes a transcritical
bifurcation at $\epsilon = 0$; for $\epsilon \ne 0$, the entire singular
set is nondegenerate. The Maslov index of $C$ is independent of
$\epsilon$ (at least for $\epsilon$ small).  For $\epsilon < 0$, $C$
encloses a line of stable fixed points with positive orientation, as determined by
\eqref{eq:sigma_j}.  For $\epsilon > 0$, $C$ encloses a line of
unstable fixed points with negative orientation.  
See Figure~\ref{fig: bifur}(a) and \ref{fig: bifur}(c).

\section{Rotationally invariant Hamiltonians in $\Rr^n$}\label{sec:rotations}

\subsection{Definitions}\label{sec: defs}
Consider $2n$-dimensional Euclidean phase space $\Rr^{2n}$
with points denoted $\zvec = (\rvec,\pvec)$.  Let $\rvec_{(j)} \in
\Rr^j$ denote the projection of $\rvec$ to its first $j$ components,
similarly $\pvec_{(j)}$ the projection of $\pvec$.  Let $\zvec_{(j)}
= (\rvec_{(j)}, \pvec_{(j})$.  Let $r_{(j)}^2 = \rvec_{(j)}\cdot
  \rvec_{(j)}$, and similarly $p_{(j)}^2$.
  
  The standard action of the rotation group $SO(n)$ on $\Rr^n$,
  $\rvec \mapsto \Rcal \cdot \rvec$, lifts to the canonical
  action $(\rvec,\pvec)\mapsto (\Rcal\cdot\rvec,\Rcal\cdot\pvec)$ on
  $\Rr^{2n}$.  The Hamiltonian generators are the
  angular momenta $L_{\alpha\beta}$ given by
\begin{equation}
  \label{eq:L_ab}
  L_{\alpha\beta} = r_\alpha p_\beta -  r_\beta p_\alpha, \quad 1\le
  \alpha,\beta \le n.
\end{equation}
Clearly $ L_{\alpha\beta} = - L_{\beta\alpha}$.  The Poisson bracket
of components of angular momenta are given by
\begin{equation}
  \label{eq:PB's of L's}
  \{L_{\alpha\beta},L_{\gamma\delta}\} =
L_{\alpha\gamma} \delta_{\beta\delta} -
L_{\alpha\delta} \delta_{\beta\gamma} -
L_{\beta\gamma} \delta_{\alpha\delta} +
L_{\beta\delta} \delta_{\alpha\gamma}.
\end{equation}
Let 
\begin{equation}
  \label{eq:L_(j)}
  L_{(j)}^2 = \half \sum_{ 1\le
  \alpha,\beta \le j}  L_{\alpha\beta}^2
 = r_{(j)}^2 p_{(j)}^2 - 
(\rvec_{(j)}\cdot\pvec_{(j)})^2
, \quad 2 < j \le n.
\end{equation}
$ L_{(n)}^2$ is the squared total angular momentum, which we
also denote by $L^2$.  More generally, $L_{(j)}^2$ is the squared
total angular momentum of the projection of $\zvec$ to $\Rr^j\times\Rr^j$.
For future reference, we note that
\begin{equation}
  \label{eq:L_j and L_j-1}
   L_{(j)}^2 =  L_{(j-1)}^2 + z_{[j]} \cdot \Qmat_{(j-1)} \cdot z_{[j]},
\end{equation}
where $z_{[j]} = (r_j,p_j)$ and 
\begin{equation}
  \label{eq:Q_j}
  \Qmat_{(j-1)} = \left(
    \begin{array}{cc}
p_{(j-1)}^2&-\rvec_{(j-1)}\cdot\pvec_{(j-1)}\\
-\rvec_{(j-1)}\cdot\pvec_{(j-1)}&r_{(j-1)}^2
    \end{array}
\right).
\end{equation}
It is easily verified that
\begin{equation}
  \label{eq:PB2s}
 \{ L_{\alpha\beta},
  L_{(j)}^2\} = 0\ \text{if $\alpha,\beta \le j$}, \quad
 \{ L_{(j)}^2,  L_{(k)}^2\} = 0.
\end{equation}

Consider a rotationally symmetric Hamiltonian $H$, characterised by
\begin{equation}
  \label{eq:H rot inv}
  \{H,L_{\alpha\beta}\} = 0, \quad 1 \le \alpha,\beta \le n.
\end{equation}
Then $H$ is integrable.  As integrals of the
motion, we may take $F_1 = H$, $F_2 = L_{12}$, and $F_j = L^2_{(j)}$
for $3 \le j \le n$.  In fact, rotationally symmetric Hamiltonians in
$\Rr^n$ are
superintegrable \cite{nekhoroshev72}; they possess (at least) $2n-2$ 
independent constants of the motion.  It would
be interesting to incorporate superintegrability into our treatment,
but here we will treat $H$ as a standard
integrable system.

\subsection{Singularities}

Singularities are points $\zvec = (\rvec,\pvec)$
for which 
\begin{equation}
  \label{eq:singularity}
  c_1 \,dH +  c_2 \, dL_{12} + \sum_{j=3}^n c_j \, dL_{(j)}^2  = 0
\end{equation}
is satisfied for nonzero  $\cvec \in \Rr^n$.
We choose a basis for the $\cvec$'s (by taking successive sets of
components to vanish).  For each basis element, we obtain
necessary and sufficient conditions for (\ref{eq:singularity}) to
hold and for the corresponding singularities to be corank-one
nondegenerate.


\subsubsection{$c_1 \ne 0$.  Spherical singularities.} 
If $c_1 \ne 0$ in (\ref{eq:singularity}), then at the singularity, the Poisson
bracket of $H$ with any rotationally invariant function $f$ must
vanish, since i) $\{f,H\} = df\cdot J\cdot dH$, ii) $dH$ may be expressed in
terms of the $dL_{\alpha\beta}$'s (cf (\ref{eq:singularity})), and
iii) $\{f,L_{\alpha\beta}\}
= df\cdot J\cdot dL_{\alpha\beta} = 0$.  There are three functionally
independent rotational invariants, eg $r^2$, $p^2$ and
$\rvec\cdot\pvec$.  The conditions $\{H,r^2\} = \{H,p^2\} =
\{H,\rvec\cdot\pvec\} = 0$ at the singularity are equivalent to
\begin{equation}
  \label{eq:H conditions}
  \rvec\cdot H_{\pvec} = 0,\quad 
  \pvec\cdot H_{\rvec} = 0,\quad
\rvec\cdot H_{\rvec} = \pvec\cdot H_{\pvec}.
\end{equation}
The conditions (\ref{eq:H conditions}) also imply a 
singularity.  To see this, note that because $H$ is rotationally
invariant, it can be expressed as a function of the invariants $r^2$,
$p^2$ and $\rvec\cdot\pvec$.  It follows that the derivatives of $H$
are of the form
\begin{equation}
  \label{eq:H gradients}
  H_{\rvec} = f \rvec + g\pvec,\quad 
H_{\pvec} = g \rvec + h\pvec,
\end{equation}
where $f$, $g$ and $h$ are functions of the invariants.
Then (\ref{eq:H conditions}) implies that
\begin{equation}
  \label{eq:dH}
  H_{\rvec} = k (p^2\rvec - (\rvec\cdot\pvec)\pvec),\quad
  H_{\pvec} = k (r^2\pvec - (\rvec\cdot\pvec)\rvec),
\end{equation}
where $k = f/p^2 = -g/(\rvec\cdot\pvec) = h/p^2$.
Therefore,
\begin{equation}
  \label{eq:form of singularity}
  dH = k dL^2,
\end{equation}
which is just (\ref{eq:singularity}) with
$c_1 = 1$, $c_n = -k$, and all other $c_j$'s equal to zero.
The $H$-orbits through such points reduce to fixed points of the
radial motion.  For this reason, we call these {\it spherical singularities}.

We determine next the condition for spherical singularities
$(\rvec,\pvec)$ to be corank-one nondegenerate.
We may assume that $\rvec$ and $\pvec$ are not both zero (as all of the
$dF_j$'s vanish at the origin).  For definiteness, let us assume that
$\rvec \ne 0$ (the treatment for $\pvec\ne 0$ is similar).  In the
neighbourhood of the singular point, we introduce local canonical
coordinates $r$, $p_r$, $L$, $\theta$ and $\Zvec$, where $p_r =
\pvec\cdot\rvec/r$ denotes the radial momentum, $L = \sqrt{L^2}$ the
magnitude of total angular momentum with conjugate variable $\theta$,
and $\Zvec$ the remaining $2n-4$ canonical coordinates.  As $H$ is
rotationally invariant, it can be expressed locally as a function of $r$,
$p_r$ and $L$, ie $H = h(r,p_r,L)$.  Then $\Kmat^2$ has only one
nonvanishing two-dimensional block, which corresponds to the $(r,p_r)$-plane.
Then
\begin{equation}
  \label{eq:radial stability}
  \half \Tr \Kmat^2 = -(h_{rr} h_{p_r p_r} - h_{r p_r}^2). 
\end{equation}
Thus, spherical singularities are nondegenerate if the corresponding
radial fixed points are nondegenerate.

\subsubsection{$(c_3,\ldots,c_n)\ne 0$, $c_1 = 0$. Axial singularities.}  
Let $m$ denote the highest index for which
$c_m \ne 0$.  We take $c_m = 1$, so that (\ref{eq:singularity}) takes
the form
\begin{equation}
  \label{eq:singularity 2}
c_2 dL_{12} +  \sum_{j=3}^{m-1} c_j  dL_{(j)}^2 + dL_{(m)}^2 = 0.
\end{equation}
The term $dL_{(m)}^2$ contains the one-forms $dr_m$ and $dp_m$,
whereas the other terms do not.  From (\ref{eq:L_j and L_j-1}), the
condition for the coefficients of $dr_m$ and $dp_m$ to vanish is that
\begin{equation}
  \label{eq:a=0,m condition}
  \Qmat_{(m-1)} \cdot z_{[m]} = 0.
\end{equation}

Let us suppose that 
\begin{equation}
  \label{eq:new equation}
  \det \Qmat_{(m-1)}\ne 0
\end{equation}
(the case $\det
\Qmat_{(m-1)}= 0$ is considered in Section \ref{sec: radial
  singularities} below).
Then (\ref{eq:a=0,m condition}) implies that
\begin{equation}
  \label{eq:rm,pm=0}
  z_{[m]} = (r_m, p_m) = 0.
\end{equation}
It is easily seen that (\ref{eq:rm,pm=0}) also implies a singularity.
For if $z_{[m]}$ vanishes, it follows from (\ref{eq:L_j and
  L_j-1}) that
\begin{equation}
  \label{eq:singularity for first condition}
  dL^2_{(m-1)} - dL^2_{(m)} = 0
\end{equation}
for $m > 3$, and for $m = 3$ that
\begin{equation}
  \label{eq:b_3}
  dL^2_3 - 2L_{12} dL_{12} = 0.
\end{equation}

For both (\ref{eq:singularity for first condition}) and (\ref{eq:b_3}),
$\Kmat$ has a single nonvanishing two-dimensional
block equal to $\Jmat\Qmat_{(m-1)}$.  It follows that
\begin{equation}
  \label{eq:Tr Kmat^2 axial}
  \half \Tr \Kmat^2 = -\det \Jmat\Qmat_{(m-1)} = -L^2_{(m)}.
\end{equation}
Thus, the nondegeneracy condition is $L^2_{(m)}\ne 0$, in which case
the singularity is elliptic.  We call singularities with $r_m = p_m =
0$ {\it $m$-axial singularities}, as the projections of their $H$-orbits
to the coordinate and momentum $m$-planes in fact lie in the
respective $(m-1)$-planes.

\subsubsection{$(c_3,\ldots,c_n)\ne 0$, $c_1 = 0$.
 Radial singularities.}  \label{sec: radial singularities}
If
\begin{equation}\label{eq:a=0 second condition}
  \det Q_{(m-1)} = L^2_{(m-1)} = 0,
\end{equation}
then (\ref{eq:a=0,m condition}) is satisfied by taking $z_{[m]}$
to be a null vector of $Q_{(m-1)}$.  It then follows from (\ref{eq:L_j
  and L_j-1}) that $L^2_{(m)} = 0$, and hence
\begin{equation}
   \label{eq:sufficient second}
   dL^2_{(m)} = 0,
\end{equation}
so that $\zvec$ is indeed singular in this case.  $L^2_{(m)} = 0$
implies that $\rvec_{(m)}$ and $\pvec_{(m)}$ are parallel; for this
reason we call such singularities {\it $m$-radial}.

$L^2_{(m)} = 0$ implies that $L^2_{(j)} = 0$ for all $2 < j < m$, and
therefore that $dL^2_{(j)} = 0$ for $2 < j < m$.  Thus, 
$m$-radial singularities have corank at least $m - 2$, and only
$3$-radial singularities can have corank one.

In fact, $3$-radial singularities are necessarily degenerate.  From
(\ref{eq:sufficient second}), we may take $c_3 = 1$ to be the only
nonzero coefficient in (\ref{eq:singularity}), so that $\Kmat =
\Jmat\cdot(L^2_{(3)})''$.  Since $L^2_{(3)} = r_{(3)}^2 p_{(3)}^2 -
(\rvec_{(3)}\cdot \pvec_{(3)})^2$, $\Kmat$ has a single nonvanishing
$6\times 6$ block given by
\begin{equation}
  \label{eq:L''_3}
\left(
  \begin{array}{cc}
-r_{(3)} p_{(3)} \Pmat_\perp&r^2_{(3)}  \Pmat_\perp\\
-p^2_{(3)}  \Pmat_\perp& r_{(3)} p_{(3)} \Pmat_\perp
  \end{array}\right),
\end{equation}
where $\Pmat_\perp$ is the projection onto the plane in $\Rr^3$
perpendicular to $\rvec_{(3)}$ and $\pvec_{(3)}$.  Since
$\Pmat_\perp^2 = \Pmat_\perp$, it follows that
\begin{equation}
  \label{eq:K^2}
  \Kmat^2 = 0.
\end{equation}

Since $3$-radial singularities are degenerate, the singularity formula
for the Maslov index cannot be applied to them.  However, as we now
show, they do not contribute to the Maslov index.  Without loss of
generality, we may assume that $n=3$.  Then a $3$-radial singularity
is of the form
\begin{equation}
  \label{eq:n = 3 radial}
  \rvec = a\nhat, \quad \pvec = b\nhat,
\end{equation}
where $\nhat$ is a unit vector in $\Rr^3$.  To ensure that the
$3$-radial singularity has corank one, we require that no other
singularity conditions are satisfied.  That is, we assume that $a$ and
$b$ do not both vanish, that (\ref{eq:H conditions}) is not satisfied
(ie, $(\rvec,\pvec)$ is not a spherical singularity), and that $\nhat$
is not  perpendicular or parallel to the $\ehat_3$, the unit vector
along the $3$-axis (ie, $(\rvec,\pvec)$ is neither a 3-axial
singularity nor
a 12-radial singularity (see below)).
Without loss of generality, we assume that $a \ne 0$ (otherwise,
reverse $a$ and $b$ in what follows).  Let $\vhat$ be a unit vector
orthogonal to $\nhat$, and let $\what = \nhat\times\vhat$.
Displacements in $\pvec$ along $\vhat$ and $\what$ with $\rvec$ held
fixed are transverse to the $3$-radial singularity.  Therefore, the
closed curve $C$, given by
\begin{equation}
  \label{eq:curve}
  \rvec(s) = a\nhat,
\quad \pvec(s) = b\nhat + \epsilon\uhat(s), 
\end{equation}
where
\begin{equation}
  \label{eq:uhat}
  \uhat(s) = \cos 2\pi s \vhat + \sin 2\pi s \what,
\end{equation}
encloses the singularity.  For sufficiently small $\epsilon$, no other
singularities are enclosed.

To calculate the Maslov index of $C$, we use the formula (\ref{eq:Maslov index formula}) explicitly.
Letting $\Avec$, $\Bvec$ and $\Cvec$ denote the three columns of
$\Mmat$, we get that
\begin{align}
  \label{eq:leading-order d's}
\Avec :=&  \frac {\partial H}{\partial \pvec} + i 
\frac {\partial H}{\partial \rvec} =
\omega \nhat + \epsilon \xi(s)\uhat(s),\nonumber \\
\Bvec :=&  \frac {\partial L_{12}}{\partial \pvec} + i 
\frac {\partial L_{12}}{\partial \rvec} = 
(a-ib)\ehat_3\times\nhat -i\epsilon \ehat_3\wedge\uhat(s),
\nonumber\\
\Cvec :=& \frac {\partial L^2}{\partial \pvec}+i \frac{\partial L^2}{\partial
   \rvec} =  2\epsilon a(a - ib)\uhat(s) + 2i\epsilon^2a \nhat,
\end{align}
where $\omega$ is a constant and $\xi(s)$ is complex and of zeroth
order in $\epsilon$.  (The expression for $\Avec$ follows from
symmetry considerations; since $H$ is rotationally symmetric, its
gradients with respect to $\rvec$ and $\pvec$ must be linear
combinations of its arguments.  The expressions for $\Bvec$ and
$\Cvec$ are obtained from straightforward calculations.)  Then
\begin{equation}
  \label{eq:det M for radial}
  |\Mmat|(\zvec(s)) = (\Cvec \times \Avec) \cdot \Bvec 
= 2 \epsilon\omega a(a-ib)((a-ib) u_3(s) + i\epsilon n_3) + O(\epsilon^3),
\end{equation}
where $u_3(s) = \ehat_3\cdot \uhat(s)$ and  $n_3 = \ehat_3\cdot \nhat$.
From (\ref{eq:det M for radial}),
\begin{equation}
  \label{eq:arg det M}
  \arg |\Mmat|(\zvec(s)) = \text{const} + \arg ((a-ib) u_3(s) +
  i\epsilon n_3) + O(\epsilon^2).
\end{equation}
The quantity $(a-ib) u_3(s) +
  i\epsilon n_3$ lies on a ray through $i\epsilon n_3$ ($n_3 \ne 0$
  by assumption), and therefore 
 has zero winding number.  The $O(\epsilon^2)$ term can be neglected.  Thus,
\begin{equation}
  \label{eq:maslov of radial}
  \mu(C) = 0.
\end{equation}

\subsubsection{$c_2 \ne 0$, $c_{j\ne 2} = 0$. (1,2)-axial singularities.}
These are singularities of the form $dL_{12} = 0$, which implies, and
is implied by,
\begin{equation}
  \label{eq:L_12 sing}
  r_1 = r_2 = p_1 = p_2 = 0.
\end{equation}
We call such singularities {\it (1,2)-axial}.  From (\ref{eq:L_12
  sing}), it is clear that (1,2)-axial singularities are also $3$-radial
singularities, and therefore have corank at least two.  
They do not contribute to the Maslov index.

\subsection{Maslov indices of rotational actions.}
Let 
\begin{equation}
  \label{eq:rot actions}
  L_{(m)} = \sqrt{ L^2_{(m)}}, \quad 3 \le m \le n.
\end{equation}
It is straightforward to verify that the Hamiltonian flow generated by
$L_{(m)}$ is a $2\pi$-periodic uniform rotation of $\rvec_{(j)}$
and $\pvec_{(j)}$ in their common plane.  That is, the orbits
generated by $L_{(m)}$ are of the form
\begin{eqnarray}
  \label{eq:(j) orbits}
  \rvec_{(m)}(s) &=& \cos s\, \rvec_{(m)} + \sin s \frac{1}{L_{(m)}} 
\left(  
r^2_{(m)}  \pvec_{(m)} -  (\rvec_{(m)}\cdot  \pvec_{(m)})  \rvec_{(m)}
\right),\nonumber\\
 \pvec_{(m)}(s) &=& \cos s \,\pvec_{(m)} - \sin s \frac{1}{L_{(m)}} 
\left(  
p^2_{(m)}  \rvec_{(m)} -  (\rvec_{(m)}\cdot  \pvec_{(m)})  \pvec_{(m)}
\right),
\end{eqnarray}
while the components $r_i$ and $p_i$ with $i > m$ are left unchanged.
The $L_{(j)}$'s, together with $L_{12}$, which generates
$2\pi$-periodic uniform rotations in the $12$-plane, constitute a set 
of $n-1$ action variables associated with rotational symmetry.  The
remaining action variable is obtained from the Hamiltonian $H$ (for
example, from the reduction of its flow to the radial phase plane).

It is straightforward to determine the Maslov indices $\mu_{12}$,
$\mu_{(3)}$, \ldots, $\mu_{(n)}$ of these rotational actions.  Let
$(\rvec,\pvec)$ be a regular point.  Consider first the angle contour
through $(\rvec,\pvec)$ generated by the flow of $L_{12}$.  This is a
$2\pi$-rotation of $\rvec_{(2)}$ and $\pvec_{(2)}$ in the $12$-plane.
The $\rvec$-orbit can be contracted to a single, nonzero point without
encountering any codimension-two singularities (in particular, it is
readily shown that the contraction can be performed keeping $r^2$,
$p^2$ and $\rvec\cdot\pvec$ fixed, so that no spherical singularities
are encountered).  The $\pvec$-orbit can then be similarly deformed
without encountering singularities.  It follows that 
\begin{equation}
  \label{eq:mu_12}
  \mu_{12} = 0.
\end{equation}

Consider next the angle contour through $(\rvec,\pvec)$ generated by
$L_{(m)}$.  From (\ref{eq:(j) orbits}), the projection of this contour
to the $(r_j,p_j)$-plane for $j \le m$ is a positively oriented ellipse about the
origin.  These $m$ ellipses can be contracted in turn to
nonzero points in their respective planes.  These deformations can be
performed keeping $r^2$, $p^2$ and $\rvec\cdot\pvec$ fixed, so that no
spherical singularities are encountered.  With each deformation up to
and including $j = 3$, a single $j$-axial singularity is encountered
(where $r_j = p_j = 0$).  Since
the ellipses are positively oriented and $j$-axial singularities are
elliptic, each contributes $+2$ to the Maslov index.  The remaining
ellipses in the $1$- and $2$-phase planes can be contracted without
encountering any singularities.   Thus,
\begin{equation}
  \label{eq:mu_{m}}
  \mu_{(m)} = 2(m-2).
\end{equation}
In particular, $\mu_{(n)}$, the Maslov index associated with the
total squared angular momentum in $n$ dimensions, is $2(n-2)$.  This
leads via (\ref{eq:EBK}) to the semiclassical quantisation condition 
$L^2 = (l + (n-2)/2)^2$, which in turn agrees with the exact
eigenvalues of the Laplacian on the $n$-sphere, $l(l+n-2)$ (see, eg,
\cite{bateman}) up to an additive $l$-independent constant.  (For $n=3$, the
semiclassical and exact eigenvalues are $(l+\half)^2$ and $l(l+1)$
respectively).

\section{Discussion}\label{sec: discussion}
For integrable systems in $\Rr^{2n}$, the Maslov index of a closed
curve in the regular component is, under certain genericity
conditions, given by a sum of contributions $\pm 2$ from the corank-one
nondegenerate singularities enclosed.  The sign depends on the
stability of the degeneracy and the orientation of the curve.  
We also obtain expressions for the transverse Liapunov
exponents of corank-one singularities.  
The fact that the index is unchanged through local bifurcations implies
relations amongst the stabilities and orientations of the singularities
involved.  For $SO(n)$-invariant systems, we recover the fact that the
Maslov indices associated with $L_{(j)}$, the magnitude of angular momentum
restricted to the first $j$ components, is $2(j-2)$.
Natural extensions of these results would include general cotangent
bundles and higher Maslov classes \cite{trofimov, suzuki}, for
which the sources should be (nondegenerate) singularities of corank greater
than one.

\bigskip
\noindent{\bf Acknowledgments} \\
We thank the referees for helpful remarks.
JAF was supported by a grant from the EPSRC.  JMR
thanks the MSRI for hospitality and support while some of this work
was carried out.

\bibliography{toda}

\begin{thebibliography}{10}

\bibitem{arnold}
V.I. Arnold.
\newblock Characteristic class entering quantization conditions.
\newblock {\em Functional Anal. Appl.}, 1:1--13, 1967.

\bibitem{arnoldcm}
V.I. Arnold.
\newblock {\em Mathematical methods of classical mechanics}.
\newblock Springer-Verlag, 2nd edition, 1989.

\bibitem{child}
M.S. Child.
\newblock Quantum states in a champagne bottle.
\newblock {\em J Phys A}, 31:657--670, 1998.

\bibitem{batescushman}
R.~Cushman and L.~Bates.
\newblock {\em Global aspects of classical integrable systems}.
\newblock Birkhäuser, 1997.

\bibitem{cdv-vn}
Y.~Colin de~Verdi\'ere and S.~Vu Ngoc.
\newblock {S}ingular {B}ohr-{S}ommerfeld rules for 2d integrable systems.
\newblock {\em Ann. Scient. \'Ec. Norm. Sup.}, 36:1--55, 2003.

\bibitem{duistermaat}
J.J. Duistermaat.
\newblock On global action-angle coordinates.
\newblock {\em Comm. Pure Appl. Math.}, 33:687--706, 1980.

\bibitem{eliasson}
L.H. Eliasson.
\newblock Normal form for {H}amiltonian systems with {P}oisson commuting
  integrals -- elliptic case.
\newblock {\em Comm. Math. Helv.}, 65:4--35, 1990.

\bibitem{bateman}
A.~Erd\'ely, W.~Magnus, F.~Oberhettinger, and F.G. Tricemi.
\newblock {\em Higher transcendental functions}, volume~2.
\newblock McGraw-Hill, 1953.

\bibitem{fomenko}
A.~Fomenko.
\newblock {\em Topological classification of integrable systems}.
\newblock Number~6 in Advances in Soviet Mathematics. AMS, 1991.

\bibitem{FR2}
J.A. Foxman and J.M. Robbins.
\newblock Singularities, {L}ax degeneracies and {M}aslov indices of the
  periodic {T}oda chain.
\newblock preprint, 2004.

\bibitem{guho}
J.~Guckenheimer and P.~Holmes.
\newblock {\em Nonlinear oscillations, dynamical systems, and bifurcations of
  vector fields}.
\newblock Springer, 1983.

\bibitem{katokhass}
A.~Katok and B.~Hasselblatt.
\newblock {\em Introduction to the Modern Theory of Dynamical Systems}.
\newblock Cambridge University Press, 1995.

\bibitem{keller}
J.B. Keller.
\newblock Corrected {B}ohr-{S}ommerfeld quantum conditions for nonseparable
  systems.
\newblock {\em Ann. Phys.}, 4:180--188, 1958.

\bibitem{litrob87}
R.G. Littlejohn and J.M. Robbins.
\newblock New way to compute {M}aslov indices.
\newblock {\em Phys.~Rev. A}, 36:2953--2961, 1987.

\bibitem{maslov}
V.P. Maslov and M.~Fedoriuk.
\newblock {\em Semiclassical approximation in quantum mechanics}.
\newblock Reidel, 1981.

\bibitem{miranda_tienzung}
E.~Miranda and N.T. Zung.
\newblock Equivariant normal form for nondegenerate singular orbits of
  integrable {H}amiltonian systems.
\newblock {\em Ann. Sci. Ecole Norm. Sup.}, 34:819--839, 2004.

\bibitem{nekhoroshev72}
N.N. Nekhoroshev.
\newblock Action-angle variables and their generalization.
\newblock {\em Trans. Moscow Math. Soc.}, 26:181--198, 1972.

\bibitem{jmr92}
J.M. Robbins.
\newblock Winding number formula for {M}aslov indices.
\newblock {\em Chaos}, 2:145--147, 1992.

\bibitem{roblit87}
J.M. Robbins and R.G. Littlejohn.
\newblock {M}aslov indices of resonant tori.
\newblock {\em Phys.~Rev. Lett.}, 58:1388--1391, 1987.

\bibitem{zhilinskii}
D.A. Sadovskií and B.I. Zhilinskii.
\newblock Monodromy, diabolic points, and angular momentum coupling.
\newblock {\em Phys. Lett. A}, 256:235--244, 1999.

\bibitem{vn2001}
Vu~Ngoc San.
\newblock Quantum monodromy and {B}ohr-{S}ommerfeld rules.
\newblock {\em Lett. Math. Phys.}, 55:205--217, 2001.

\bibitem{suzuki}
H.~Suzuki.
\newblock Residue classes of {L}agrangian subbundles and {M}aslov classes.
\newblock {\em Trans. Amer. Math. Soc.}, 347:189--202, 1995.

\bibitem{toth}
J.A. Toth and S.~Zelditch.
\newblock {$L^p$} norms of eigenfunctions in the completely integrable case.
\newblock {\em Ann. Henri Poincar\'e}, 4:343--368, 2003.

\bibitem{trofimov}
V.V. Trofimov.
\newblock Generalized {M}aslov classes on the path space of a symplectic
  manifold.
\newblock {\em Proc. Steklov Institute of Mathematics}, 205:157--179, 1995.

\bibitem{vey}
J.~Vey.
\newblock Sur certains syst\`emes dynamiques s\'eparables.
\newblock {\em Amer. J. Math.}, 100:591--614, 1978.

\bibitem{tienzung}
N.T. Zung.
\newblock Symplectic topology of integrable {H}amiltonian systems, {I}:
  {A}rnold-{L}iouville with singularities.
\newblock {\em Compositio Math.}, 101:179--215, 1996.

\end{thebibliography}

\end{document}